\documentclass[a4paper,12pt]{article}
\usepackage{amsfonts,amsthm,amsmath,amssymb,graphicx,color,hyperref,youngtab}
\usepackage[bulletsep]{collref}
\usepackage[lmargin=3.0cm, rmargin=1.5cm,tmargin=2.50cm,bmargin=2.50cm]{
geometry}
\usepackage{mathrsfs}
\usepackage{float}
\usepackage{ulem}
\usepackage{graphicx}

\newcommand{\be}{\begin{equation}}
\newcommand{\ee}{\end{equation}}
\newcommand{\bea}{\begin{eqnarray}}
\newcommand{\eea}{\end{eqnarray}}

\newcommand{\nn}{{\nonumber\\}}
\newcommand{\Tr}{\text{Tr}}

\newcommand\beq{\begin{equation}}
\newcommand\eeq{\end{equation}}

\begin{document}


\def\gap#1{\vspace{#1 ex}}
\def\be{\begin{equation}}
\def\ee{\end{equation}}
\def\bal{\begin{array}{l}}
\def\ba#1{\begin{array}{#1}}  
\def\ea{\end{array}}
\def\bea{\begin{eqnarray}}
\def\eea{\end{eqnarray}}
\def\beas{\begin{eqnarray*}}
\def\eeas{\end{eqnarray*}}
\def\del{\partial}
\def\eq#1{(\ref{#1})}
\def\fig#1{Fig \ref{#1}} 
\def\re#1{{\bf #1}}
\def\bull{$\bullet$}
\def\nn{\\\nonumber}
\def\ub{\underbar}
\def\nl{\hfill\break}
\def\ni{\noindent}
\def\bibi{\bibitem}
\def\ket{\rangle}
\def\bra{\langle}
\def\vev#1{\langle #1 \rangle} 
\def\lsim{\stackrel{<}{\sim}}
\def\gsim{\stackrel{>}{\sim}}
\def\mattwo#1#2#3#4{\left(
\begin{array}{cc}#1&#2\\#3&#4\end{array}\right)} 
\def\tgen#1{T^{#1}}
\def\half{\frac12}
\def\floor#1{{\lfloor #1 \rfloor}}
\def\ceil#1{{\lceil #1 \rceil}}

\def\mysec#1{\gap1\ni{\bf #1}\gap1}
\def\mycap#1{\begin{quote}{\footnotesize #1}\end{quote}}
\def\ubsec#1{\gap1\ni\underbar{#1}\gap1}

\def\Om{\Omega}
\def\a{\alpha}
\def\b{\beta}
\def\l{\lambda}
\def\m{\mu}
\def\n{\nu}
\def\om{\omega}

\def\lan{\langle}
\def\ran{\rangle}

\def\bit{\begin{item}}
\def\eit{\end{item}}
\def\benu{\begin{enumerate}}
\def\eenu{\end{enumerate}}
\def\eps{\epsilon}


\def\bT{{\bar T}}
\def\bL{{\bar L}}
\def\lt{{\tilde \lambda}}
\def\vt{{\tilde v}}
\def\wt{{\tilde w}}
\def\omt{{\tilde\omega}}
\def\ut{{\tilde u}}
\def\bP{{\bf P}}
\def\bQ{{\bf Q}}
\def\Jt{{\tilde J}}



\rightline{TIFR/TH/14-14}
\vspace{1.2truecm}

\vspace{1pt}


{\LARGE{
\begin{center}{\bf The {\it inside out}s of
AdS$_3$/CFT$_2$: Exact AdS wormholes with
entangled CFT duals}
\end{center}
}}

\vskip.9cm

\thispagestyle{empty} \centerline{\large \bf  
Gautam Mandal,\footnote{mandal@theory.tifr.res.in} Ritam Sinha,\footnote{ritam@theory.tifr.res.in} and Nilakash Sorokhaibam\footnote{nilakashs@theory.tifr.res.in}}
    
\vspace{.8cm} 
\centerline{{\it Department of Theoretical Physics}}
\centerline{{\it Tata Institute of Fundamental Research, Mumbai
    400005, India.} }


\gap7

\centerline{\today}

\gap3

\thispagestyle{empty}

\gap6

\centerline{\bf Abstract}
\vskip.5cm 

We present the complete family of solutions of 3D gravity
($\Lambda<0$) with {\it two} asymptotically AdS exterior regions. The
solutions are constructed from data at the two boundaries, which
correspond to two independent and arbitrary stress tensors $T_R, \bar
T_R$, and $T_L, \bar T_L$. The two exteriors are smoothly joined on to
an interior region through a regular horizon. We find CFT duals of
these geometries which are entangled states of two CFT's.  We compute
correlators between general operators at the two boundaries and find
perfect agreement between CFT and bulk calculations. We calculate and
match the CFT entanglement entropy (EE) with the holographic EE which
involves geodesics passing through the wormhole. We also compute a
holographic, non-equilibrium entropy for the CFT using properties of
the regular horizon. The construction of the bulk solutions here uses
an exact version of Brown-Henneaux type diffeomorphisms which are
asymptotically nontrivial and transform the CFT states by two
independent unitary operators on the two sides.  Our solutions provide
an infinite family of explicit examples of the ER=EPR relation of
Maldacena and Susskind \cite{Maldacena:2013xja}.

\setcounter{page}{0} \setcounter{tocdepth}{2}

\newpage

\tableofcontents

\section{\label{sec:intro}Introduction and Summary}

It has been a matter of lively debate whether the standard description
of a large black hole with a smooth horizon is quantum mechanically
consistent, and is, in fact, consistent with AdS/CFT. While the
firewall hypothesis \cite{Almheiri:2012rt, Almheiri:2013hfa}
\footnote{See also \cite{Braunstein:2009my}.} argues
against the validity of the standard description, Maldacena and
Susskind \cite{Maldacena:2013xja} have suggested that the region inside
the horizon is a geometric representation of quantum mechanical
entanglement. Both the above proposals, and related issues, are
discussed in a number of papers; for a partial list, related to the
discussion in this paper, see
\cite{Almheiri:2012rt,Almheiri:2013hfa,Shenker:2013pqa,
  VanRaamsdonk:2013sza,Marolf:2013dba,
  Papadodimas:2013jku,Papadodimas:2013wnh,Shenker:2013yza,
  Avery:2013bea,Balasubramanian:2014gla}. The proposal of
\cite{Maldacena:2013xja}, summarized by the symbolic equation ER =
EPR, \footnote{Einstein-Rosen (wormhole) = Einstein-Podolsky-Rosen
  (entangled state).} is illustrated by the eternal black hole
geometry which is dual to the thermofield state
\cite{Maldacena:2001kr}.\footnote{See \cite{Hartman:2013qma} for an
  AdS/CFT check on the dynamical entanglement entropy which involves
  the wormhole region, and \cite{Caputa:2013eka} for generalization to
  include angular momentum and charge.} It has been argued in several
papers (see, e.g., \cite{Marolf:2013dba, Balasubramanian:2014gla})
that although the proposal holds for this illustrative case, it does
not hold in general. One of the objectives of the present work is to
explicitly construct a general class of two-sided
geometries \footnote{By {\it two-sided}, we mean geometries which have
  two asymptotically AdS regions.} which represent entangled CFT's.

A useful approach to construct the geometric dual to a CFT state is by
using a Fefferman-Graham (FG) expansion, with boundary data provided
by the CFT state. To begin with, let us consider the case of a single
CFT.  Since we are primarily interested in the metric, let us focus,
for simplicity, on states in which only the stress tensor is excited.
The dual geometry would then be given by the solution to the
appropriate Einstein equations subject to the boundary data provided
by the stress tensor. This approach has been particularly fruitful in
the context of the AdS$_3$/CFT$_2$ duality where the Fefferman-Graham
expansion has been shown, for pure gravity, to terminate
\cite{Banados:1998gg} , yielding the following exact
metric \footnote{In \eq{banados}, $x_\pm= t\pm x$, with $x \in
  {\mathbb R}$.  For $L, \bar L$ constant, this corresponds to the BTZ
  black string.}
\begin{align}
ds^2 =   \frac{dz^2}{z^2}  - dx_+ dx_-\left(\frac1{z^2} + 
z^2 \frac{L(x_+) \bL(x_-)}{16} \right) +\frac14  \left(L(x_+) dx_+^2  + 
\bL(x_-) dx_-^2 \right)
\label{banados}
\end{align}
The boundary data ($z\to 0$) is represented by the following
holographic stress tensors (we choose $-\Lambda =1/\ell^2=1$)
\begin{align}
8\pi G_3 T_{++}(x_+)=  \frac{L(x_+)}4, \,  
8\pi G_3 T_{--}(x_-)=  \frac{\bL(x_-)}4
\label{banados-stress}
\end{align} 
The above metric becomes singular at the horizon
\be 
z= z_0 \equiv 2\left(L(x_+) \bL(x_-)\right)^{-1/4},
\label{banados-horizon}
\ee and therefore the metric \eq{banados}, describes only an exterior
geometry. \footnote{The inverse metric $g^{MN}$ blows up at the
  horizon, as in case of Schwarzschild geometry. However, unlike
  there, here the other region $z>z_0$ does {\it not} represent the
  region behind the horizon; rather it gives a second coordinatization
  of the exterior region again.  In this paper, we will use a
  different set of coordinate systems to probe the interior and a
  second exterior region.}

How does one carry out such a construction with two boundaries, with
two sets of boundary data? Indeed, it is not even clear, {\it a
  priori}, whether simultaneously specifying two independent pieces of
boundary data can always lead to a consistent solution in the bulk
(this question has been raised in several recent papers, e.g. 
see \cite{VanRaamsdonk:2013sza}). A possible approach to this problem
is suggested by the fact that the eternal BTZ solution, which contains
\eq{banados} with constant stress tensors, admits a maximal extension
with two exteriors, which are joined to an interior region across a
smooth horizon. The maximal extension is constructed by transforming,
e.g., to various Eddington-Finkelstein (EF) coordinate patches
(described in Appendix \ref{charts}). A naive generalization of such a
procedure in case of variable $L, \bL$, of transforming the metric
\eq{banados} to EF type coordinates, does not seem to work since it
leads to a complex metric in the interior region \footnote{Such a
  coordinate transformation has been discussed in \cite{Gupta:2008th}
  in an asymptotic series near the boundary.}.  A second approach
could be to solve Einstein's equations, by using the constant $L, \bL$
(eternal BTZ) solution as a starting point and, incorporate the effect
of variable $L, \bL$ perturbatively, either in a derivative expansion
or an amplitude expansion.  While this method may indeed work, at the
face of it, it is far from clear how the variation in $L, \bL$ can be
chosen to be different at the two boundaries.

In this paper, we will use the method of solution generating
diffeomorphisms (SGD). In gauge theory terms, these are asymptotically
nontrivial gauge transformations which correspond to global charge
rotations; the use of these objects was introduced in
\cite{Regge:1974zd,Wadia:1979yu, Gervais:1976ec}, and used crucially
by Brown and Henneaux\cite{Brown:1986nw} to generate `Virasoro
charges' through asymptotically nontrivial SGDs that reduced at the
AdS boundary to conformal transformations.  (We discuss these in more
detail in Section \ref{behind}). Brown and Henneaux had discussed only
the asymptotic form of the SGDs. We apply two independent, exact
Brown-Henneaux SGDs \footnote{\label{ftnt:gauge} It has been shown by
  Roberts \cite{Roberts:2012aq} that the exterior metric \eq{banados}
  can be obtained by an exact Brown-Henneaux type diffeomorphism
  applied to the Poincare metric. See Appendix \ref{roberts} for a
  discussion on this and a different, new, transformation which is
  closer to the ones we use in this paper.} to different coordinate
patches of the eternal BTZ geometry, yielding a black hole spacetime
with two completely general stress tensors on the two boundaries.  In
other words, our strategy for solving the boundary value problem can
be summarized as: given arbitrary boundary data in terms of stress
tensors $T_R, \bar T_R$, and $T_L, \bar T_L$, we (i) find the two
specific sets of conformal transformations (which we are going to call
$G_+, G_-$ and $H_+, H_-$) which, when acting on a constant stress
tensor, gives rise to these stress tensors, (ii) find the SGD's which
reduce to these conformal transformations and (iii) apply the SGD's to
the eternal BTZ metric.

\gap1

\ni{\it This solves the boundary value problem we
  posed above.} 

\gap1

The results in this paper are organized as follows:

(1) \underbar{The new solutions}: In Section \ref{behind} we describe
the explicit {\it solution generating diffeomorphisms} (SGDs) and
construct the resulting two-sided black hole geometries. The
diffeomorphisms reduce to conformal transformations at each boundary,
parameterized by functions $G_\pm$ on the right and $H_\pm$ on the
left. The SGD parameterized by $G_\pm$ is applied to the
Eddington-Finkelstein coordinate chart EF1 (which covers the right
exterior and the black hole interior, see Figure \ref{fig-ef1234}) and
to EF4 (right exterior + white hole interior), whereas the SGD
parameterized by $H_\pm$ is applied to the Eddington-Finkelstein
coordinate chart EF2 (left exterior + black hole interior) and to EF3
(left exterior + white hole interior). To cover the entire spacetime
we also use a Kruskal chart K5 which covers an open neighbourhood of
the bifurcate Killing horizon; here we leave the original Kruskal
metric unaltered.  The effect of the above SGDs is that we have a
description of different metric tensors in different charts. In
Section \ref{full-metric} we show that all these can be pieced
together to give a single (pseudo-)Riemannian manifold; we prove this
by showing that in the pairwise overlap of any two charts $N_1 \cap
N_2$ the different metrics constructed above differ only by a trivial
diffeomorphism (see the definition \ref{def-nontrivial}); the
full metric, specified with the help of the various 
charts, is schematically represented in Figure \ref{fig-nontrivial}. 
An important
manifestation of the asymptotic nontriviality of the SGDs is to move
and warp the infra-red regulator surface (see Figure \ref{fig-ef});
the change in the boundary properties, as found in later sections, can
be directly attributed to this.

The new spacetime so constructed inherits the original causal
structure, with the event horizon, the bifurcation surface, and the
two exterior and interior regions (see also footnotes \ref{inner-a}
and \ref{inner}). The horizon is, therefore, regular by
construction. In the new EF coordinates (the {\it tilded} coordinates)
the horizon consists of smoothly undulating surfaces (see Fig
\ref{fig-horizons}).

(2) \underbar{The CFT duals}: In section \ref{CFT-dual} we use the
observation that the SGDs reduce asymptotically to conformal
transformations to assert that the CFT duals to our geometries are
given by unitary transformations $U_L \otimes U_R$ to the thermofield
double state. Since the unitary transformations implement conformal
transformations, AdS/CFT implies that CFT correlators in the
transformed state are holographically computed by the new spacetime
geometry. We posit this as a test of the proposed AdS/CFT
correspondence.

(3) \underbar{The AdS/CFT checks}: In section \ref{sec-T} we carry out
this test for the stress tensor. We compute the holographic stress
tensor \cite{Balasubramanian:1999re,Skenderis:1999nb} in the new
geometry and show that it exactly matches with the expectation value
of the conformally transformed (including the Schwarzian derivative)
stress tensor in the thermofield double state.  In section \ref{2-pt}
we compare AdS and CFT results for both $\lan O_L O_R \ran$ and $\lan
O_R O_R \ran$ types of correlators. The holographic two-point function
is found by computing geodesic lengths in the new geometries and we
find that it correctly matches with the two-point function of
transformed operators. This can be regarded as an evidence for the
ER=EPR relation in the presence of probes.

(4) \underbar{Entanglement entropy}: As a further check, in section
\ref{entang} we apply the above result for two-point functions to show
that the entanglement entropy EE in CFT matches the holographic EE
\cite{Ryu:2006bv,Hubeny:2007xt} including when the Ryu-Takayanagi
geodesic passes through the wormhole. This constitutes a direct proof
of the ER=EPR conjecture for the entire class of geometries constructed in this
paper.  We work out the dynamical entanglement entropy in an example
(see fig \ref{HEEplot}).

(5) \underbar{Holographic entropy from horizon}: In section
\ref{sec-S}, we make crucial use of the existence of smooth horizons
on both sides to compute a holographic entropy along the lines of 
\cite{Bhattacharyya:2008xc}. We are able to compute the entropy in the
CFT by using the Cardy formula and an adiabatic limit (which allows
the use of the `instantaneous' energy eigenvalues to compute
degeneracies); the holographic entropy agrees with this. The entropy
turns out to be divergenceless, reflecting the dissipationless nature
of 2D CFT. There is, however, a nontrivial local flow of entropy (see
fig \ref{entropy}).

(6) \underbar{ER=EPR}: In Section \ref{discuss} we discuss some
implications of our solutions {\it vis-a-vis} the ER=EPR relation of
Maldacena and Susskind \cite{Maldacena:2013xja}. Our solutions
establish an infinite family of quantum states entangling two CFTs
which are represented in the bulk by wormhole geometries. We show, in
particular, that out of a given set of quantum states we consider, all
characterized by the same energy, there are states with low
entanglement entropies, which nevertheless are still represented by
wormhole geometries; this is in keeping with the picture of geometric
entanglement suggested in \cite{Maldacena:2013xja}.

\section{The solutions\label{behind}}

In this section we obtain the new solutions by carrying out the
procedure outlined in the Introduction.  As explained in Section
\ref{charts}, for constant $L, \bL$, the metric \eq{banados}
represents a BTZ black hole of constant mass and angular momentum
\eq{btz-mass-ang}. In that case, one can construct EF coordinates (see
Section \ref{charts}) to extend the spacetime to include the region
behind the horizon and a second exterior. We will, in fact, use five
charts to cover the extended geometry (see Fig \ref{fig-ef1234}).

\gap{-3}
\begin{figure}[H]
\centerline{\includegraphics[width=550pt,height=150pt]{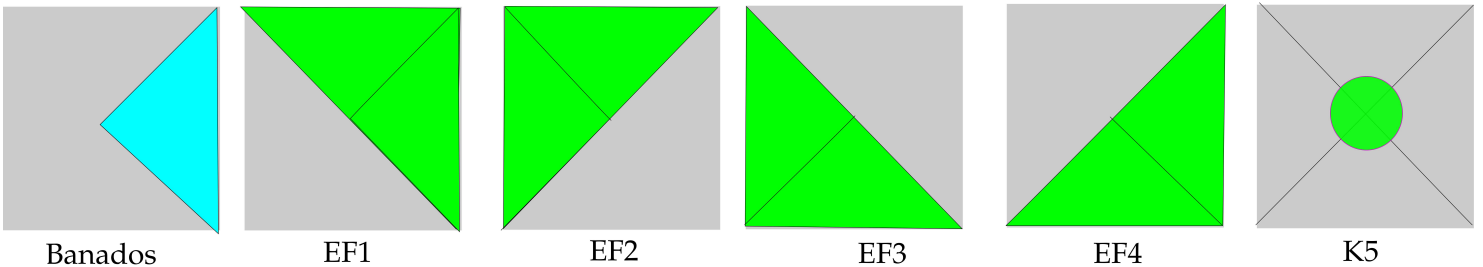}}
\caption{The (green parts of) the five figures on the right depict the
  five coordinate charts used in this paper to cover the eternal BTZ
  solution.\protect\footnotemark The coordinate chart K5 is needed
  to cover the ``bifurcation surface'' where the past and future
  horizons meet (it is a point in the Penrose diagram). The leftmost
  diagram (in blue) represents the coordinate chart used in
  \eq{banados}. Each of the coordinate charts is shown, for facility
  of comparison, within a Penrose diagram where the parts not within
  the chart are shown in gray.}
\label{fig-ef1234}
\end{figure}
\footnotetext{\label{inner-a} This is the entire geometry for
    the non-spinning BTZ; for spinning BTZ solutions, we do not
    attempt to cover the region beyond the inner horizon, since in
    this paper we are interested in the asymptotic properties in the
    two exteriors mentioned above.  See also footnote \ref{inner}.}

\subsection{\label{eternal}The eternal BTZ geometry}

We will now briefly review some properties of the eternal BTZ
geometry.  The maximal extension of the eternal BTZ geometry, starting
from \eq{banados} is described in detail in Section \ref{charts}.  We
will briefly reproduce some of the formulae relevant to the
coordinate system (``EF1'') describing the right exterior and the
interior.  The EF1 coordinates are obtained from the coordinates of
\eq{banados} by the transformations
\begin{align}
\frac{z}{z_0} &= \sqrt{\frac1{\l_0} \left( \l - \sqrt{\l^2- \l_0^2} 
\right)}
\nn
x_+ &= v- \frac1{2\sqrt L}\ln\left(\frac{\l -\l_0}{\l + \l_0}\right),\;
x_- = w- \frac1{2\sqrt{\bL}}\ln\left(\frac{\l -\l_0}{\l + \l_0}\right)
\end{align}
The metric, in these coordinates, becomes
\begin{eqnarray}
 ds^2=\frac{d\lambda^2}{4(\lambda+\lambda_0)^2}+\frac{L}{4}dv^2+\frac{\bL}{4}dw^2-\lambda\ dv dw
 +\frac{\sqrt{L}}{2(\lambda+\lambda_0)}dv d\lambda+\frac{\sqrt{\overline{L}}}{2(\lambda+\lambda_0)}dw d\lambda
\label{EF}
\end{eqnarray}
The event horizon $\l_H$, the inner horizon
$\l_i$, and the singularity $\l_s$ are at 
\begin{align}
\l_H &= \l_0 \equiv \frac{\sqrt{L\bL}}2, \; \l_i = - \l_0,\;
\l_s = - \frac14 (L +\bL)
\label{lam-horizon}
\end{align}
Note that for BTZ black holes without angular momentum $\bL=L$
and $\l_i = \l_s$. The location of the event horizon corresponds to
\eq{banados-horizon}.

In order to regulate IR divergences coming from $\l \to \infty$,
we define a cut-off surface $\Sigma_B$ at a constant large 
$\l= \l_{ir}$; the metric \eq{EF} on $\Sigma_B$ turns out
to be 
\be
\l=\l_{ir}= 1/\eps^2 \Rightarrow  
ds^2|_{\Sigma_B} =  -(1/\eps^2)\ dv\ dw (1+
O(\eps^2))
\label{lam-ir}
\ee 
By the usual AdS/CFT correspondence the leading
term defines the boundary metric (see Section \ref{uv-ir})
\be
ds^2_{bdry}= - dv\ dw
\label{bdry-lam}
\ee
The subleading term in the metric corresponds to the normalizable
metric fluctuation, which gives the expectation value of the stress tensor;
this is the holographic stress tensor \cite{Balasubramanian:1999re}, and 
is given here by 
\begin{align}
8\pi G_3 T_{vv}(x_+)=  \frac{L}4, \,  
8\pi G_3 T_{ww}(x_-)=  \frac{\bL}4
\label{lam-stress}
\end{align} 

It is easy to see that we will get the same boundary metric and stress
tensor from an analysis of the coordinate chart EF4. It is also
straightforward to derive similar results for the left exterior
(which represent a state with the same mass and angular momentum)
using EF2 and EF3.  

\subsection{\label{SGD}Solution generating diffeomorphisms (SGD)}

We will now proceed to construct new solutions with arbitrary boundary
data at the two boundaries (represented by two arbitrary holographic
stress tensors $T_{R, \mu\nu}(x)$ and $T_{L,\mu\nu}(x)$) by applying the
method of solution generating diffeomorphisms to the above geometry,
as explained in the introduction.

The solution generating diffeomorphisms can be described as
follows. Suppose we start with a certain metric $g_{MN}(x) dx^M
dx^N$ \footnote{Notation: $x^M=\{ \l, x^\mu \}$,\; $x^\mu = \{v,w\}$.}
in a certain coordinate chart ${\cal U}_P$ containing a point P.  The
new metric $\tilde g_{MN}$, in this coordinate chart, is given in
terms of a diffeomorphism (active coordinate transformation) $f:{\tilde x}^M=
{\tilde x}^M(x)$, by the definition
\be
g\to \tilde g\equiv f^*g: \quad
\tilde g_{MN}(\tilde x) \equiv  \frac{\del x^P}{\del {\tilde x}^M}  
\frac{\del x^Q}{\del {\tilde x}^N} g_{PQ}(x)
\label{diffeo-gen}
\ee
In the above, $f^*g$ is a standard mathematical notation for
the pullback of the metric $g$ under the diffeomorphism $f$.  
For diffeomorphisms differing infinitesimally from the identity
map: $\tilde x^M = x^M - \xi^M(x)$, we, of course, have the
familiar relation
\be
\delta g_{MN}(x)= D_M \xi_N + D_N \xi_M
\label{killing}
\ee
Normally, a diffeomorphism is considered giving rise to a physically
indistinguishable solution; this, however, is not true  
when the diffeomorphism is non-trivial at infinity (this
is explained in more detail in Section \ref{nontrivial}). 

As explained in Section \ref{charts}, we use five charts to cover the entire
eternal BTZ geometry (see Fig \ref{fig-ef1234}). These charts are labelled as
EF1, EF2, EF3, EF4 and K5. 
We use a nontrivial diffeomorphism in each of EF1, EF2, EF3 and EF4, which
overlap with the boundary and the identity transformation in the
Kruskal patch K5.

\subsubsection{The metric in the coordinate chart EF1}

The diffeomorphism in the EF1 coordinate chart is given by 
\be 
\l=\frac{\lt}{G_+'(\vt) G_-'(\wt)}, \; v= G_+(\vt), \;
w= G_-(\wt)
\label{diffeo}
\ee
The new metric $\tilde g_{MN}$, written in terms of 
${\tilde x}^M = (\lt, \vt, \wt)$, is 
\begin{align}
\tilde g_{MN}(\tilde x) d{\tilde x}^M d{\tilde x}^N
\equiv   
ds^2&=\frac{1}{B^2}
 \left[
  d\lt^2 +
 A_+^2 d\vt^2+
 A_-^2 d\wt^2+
 2A_+ d\vt d\lt +
 2A_- d\wt d\lt 
 \right. 
 \nonumber\\
 &\kern-30pt - \left.
 \lt \bigg(
 B^2+2\bigg(A_+\frac{G''_-(\wt)}{G_-'(\wt)}+A_-\frac{G_+''(\vt)}{G_+'(\vt)}+\lt \frac{G_+''(\vt)G_-''(\wt)}{G_+'(\vt)G_-'(\wt)}
 \bigg)\bigg)d\vt d\wt
 \right]
\label{newmetric}
\end{align}
where
\begin{eqnarray}
A_+=\sqrt{L} G_+'(\vt)(\lt+\lt_0)-\lt \frac{G_+''(\vt)}{G_+'(\vt)}, 
\;
 A_-=\sqrt{\bL} G_-'(\wt)(\lt+\lt_0)-\lt \frac{G_-''(\wt)}{G_-'(\wt)}, \;
 B=2(\lt+\lt_0)\nonumber
\end{eqnarray}
For infinitesimal transformations $G_\pm(x) \equiv x + \eps_\pm(x)$,
this amounts to an asymptotically nontrivial 
diffeomorphism $\xi^M$ (see \eq{killing})\footnote{The subscript
in $\xi_1^M$ refers to the chart EF1.}
\be
\xi_1^v= \eps_+(v),\; \xi_1^w= \eps_-(w), \; \xi_1^\l
=- \l \left( \eps_+'(v) +  \eps_-'(w) \right)
\label{killing-a}
\ee
The behaviour of the metric \eq{newmetric} at a constant
large $\l$ surface is given by
\be
ds^2= - \lt \ d\vt d\wt\ (1 +  O(1/\lt))
\label{leading-lamt}
\ee
This, by following arguments similar to the previous case
(see Section \ref{eternal}), identifies the IR cutoff
surface as 
\be
\lt_{ir}= (1/\eps^2) 
\label{lamt-ir}
\ee
and the boundary metric as 
\begin{align}
ds^2_{bdry} &= - d\vt d\wt\
\label{boundary-metric}
\end{align}
The subleading term in
\eq{leading-lamt}, as explored in Section \ref{sec-T}, gives the
holographic stress tensor. We will see there that the subleading term
depends on the SGD functions $G_\pm$; this feature is what makes the
SGD's asymptotically \emph{nontrivial} (see Section \ref{nontrivial} for a
more detailed discussion on this).

In terms of the old $\l$-coordinate, the surface \eq{lamt-ir} is
\be
\l= 1/(\eps^2 G_+'(\vt) G_-'(\wt)) 
\label{lamt-lam-ir}
\ee
Note that this surface is different from \eq{lam-ir}, and is
nontrivially warped, as in Figure \ref{fig-ef}.  This is another
manifestation of the asymptotic non-triviality of the diffeomorphism
\eq{diffeo}, which is responsible for nontrivial transformation of
bulk quantities, such as geodesic lengths.

\begin{figure}[H]
\centerline{\includegraphics[scale=.20]{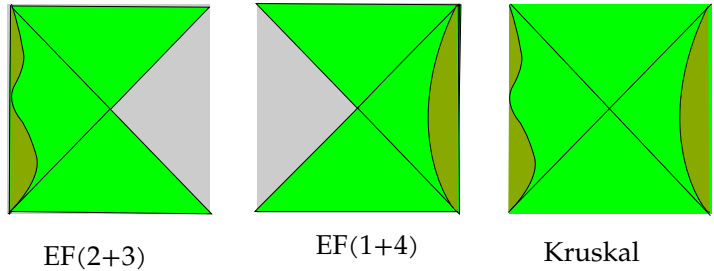}}
\caption{This figure shows the IR cut-off \eq{lamt-ir} in the new
  geometries. The effect of the SGDs, in the old (un-tilded)
  coordinates, is to deform the IR cut-off surfaces. The surface
  deformation on the right exterior is given by the change from
  \eq{lam-ir} to \eq{lamt-lam-ir}; there is a similar surface
  deformation on the left exterior.}
\label{fig-ef}
\end{figure}

We note that the leading large $\lt$ behaviour of \eq{newmetric} is
that of  AdS$_3$
\be
ds^2 =  \frac{d\lt^2}{4 \lt^2}  - \lt\ d\vt\ d\wt + ...
\label{asym-ads3}
\ee 
As mentioned before, and will be explored in detail in
Section \ref{sec-T}, the subleading terms, represented by 
the ellipsis $...$, are nontrivially different from that of AdS$_3$.
\subsubsection{The metric in the coordinate chart EF2}

The diffeomorphism (SGD) used in the coordinate chart EF2
(see Fig \ref{fig-ef1234}), which is independent of
the one above used  in EF1, is given by
\be
\l_1=\frac{\lt_1}{H_+'(\ut) H_-'(\omt)},\; u=H_+(\ut),\;\om=H_-(\vt)
\ee
which leads to the metric 
\begin{align}
 ds^2&=\frac{1}{B^2}
 \left[
 d\lt_1^2+
 A_+^2 d\ut^2+
 A_-^2 d\omt^2
 -2 A_+ d\ut d\lt_1
 -2 A_-d\omt d\lt_1
 \right. 
 \nonumber\\
 & - \left.
\lt_1 \bigg(
 B^2-2\bigg(
 A_+\frac{H_-''(\omt)}{H_-'(\omt)}+A_-\frac{H_+''(\ut)}{H_+'(\ut)}-\lt_1\frac{H_+''(\ut)H_-''(\omt)}{H_+'(\ut)H_-'(\omt)}
 \bigg)\bigg)
 d\omt d\ut
 \right]
\label{newmetric-2}
\end{align}
where
\begin{eqnarray}
 A_+=\sqrt{L} H_+'(\ut)(\lt_1+\lt_0)+ \lt_1 \frac{H_+''(\ut)}{H_+'(\ut)},\; 
 A_-=\sqrt{\bL} H_-'(\omt)(\lt_1+\lt_0)+ \lt_1 \frac{H_-''(\omt)}{H_-'(\omt)},\;
 B=2(\lt_1+\lt_0)\nonumber
\end{eqnarray}
For infinitesimal transformations $H_\pm(x)= x + \varepsilon_\pm(x)$, this
implies a diffeomorphism  $\xi_2^M$ where
\be
\xi_2^u=- \varepsilon_+(u),\; \xi_2^\om=- 
\varepsilon_-(\om), \; \xi_2^\l
= -\l \left( \varepsilon_+'(u) +  \varepsilon_-'(\om) \right)
\label{killing-b}
\ee
Note, once again, the asymptotic nontriviality of the above diffeomorphism.

\subsection{\label{full-metric}The full metric}

In a manner similar to the above, we apply the SGD characterized by
$G_\pm$ on EF4 (which shares the right exterior with EF1, see Appendix
\ref{EFs}): and the SGD characterized by $H_\pm$ on EF3 (which shares
the left exterior with EF2):
\begin{align}
\hbox{EF4}:\kern10pt &\l=\frac{\lt}{G_+'(\ut_1) G_-'(\omt_1)}, \;
u_1=G_+(\ut_1),\;\om_1=G_-(\omt_1)
\nonumber\\
&\kern20pt \hbox{infinitesimally}~~
\left(\xi_4^\l,\xi_4^{u_1}, \xi_4^{\om_1}\right)
= \left(-\l (\eps_+'(u_1) + \eps_-'(\om_1)), \eps_+(u_1), \eps_-(\om_1)
\right)
 \nonumber\\
\hbox{EF3}:\kern10pt &\l=\frac{\lt_1}{H_+'(\vt_1) H_-'(\wt_1)}, \;
v_1=H_+(\vt_1),\;w_1=H_-(\wt_1)
\nonumber\\
&\kern20pt \hbox{infinitesimally}~~
\left(\xi_4^\l,\xi_4^{v_1}, \xi_4^{w_1}\right)
= \left(-\l (\varepsilon_+'(v_1) + 
\varepsilon_-'(w_1)), \varepsilon_+(v_1), \varepsilon_-(w_1)
\right)
\end{align}

The infinitesimal transformations are similar to those in
eqs. \eq{killing-a} and \eq{killing-b}.  As mentioned above, we use
the identity diffeomorphism of Kruskal patch K5 (with $\xi_5^M=0$).
The expressions for the metric in various coordinate charts
are given in \eq{newmetric}, \eq{newmetric-2}, \eq{newmetric-3},
\eq{newmetric-4} and \eq{Kruskal}.

We will now show that the five different metrics in the five
coordinate charts define a single metric in the entire spacetime.  To
see this, note that although the SGD's applied on the five charts
are different, (equivalently, for infinitesimal transformations, the
diffeomorphisms $\xi_i^M$ in the five charts differ from each other),
they satisfy the following sufficient criteria:

\begin{enumerate}

\item [(i)] At both the right (and left) exterior boundary, the
  diffeomorphisms coincide. For example, in case of the right exterior (see
  \eq{rt-exterior}), as $\l \to \infty$, $u_1 \to v $, $\om_1 \to w$.
  Hence $\ut_1 = G_+^{-1}(u_1) \to G_+^{-1}(v) = \vt$. In other words,
  for infinitesimal transformations $\xi_4^M(P)\to \xi_1^M(P)$ for a
  given point $P$ with $\l \to \infty$. This implies that the metric
  \eq{newmetric} coincides at the right boundary with the similar
  metric\eq{newmetric-3} obtained by applying the $G_\pm$ transformations
  on the coordinate chart  EF4. Similarly, the metric \eq{newmetric-2} obtained by the
  $H_\pm$ transformations in EF2 and the similar metric \eq{newmetric-4} obtained by the
  $H_\pm$ transformations in EF3 coincide at the left exterior boundary. 

\item [(ii)] Away from the boundary, the metrics obtained in the
  various EF coordinate charts differ from each other only by trivial
  diffeomorphisms which become the identity transformation at
  infinity. Since the physical content of each of these metrics is
  represented only by the boundary data, the above point (i) ensures
  that all the different metrics represent the same single spacetime
  metric in different charts (see Figure \ref{fig-nontrivial}).

\item[(iii)] It is clear that the SGDs lead to a {\it smooth metric}
  in each chart, provided $G_\pm(x), H_\pm(x)$ are differentiable and
  invertible functions. In the rest of the paper, we will only
 consider such functions. It can be verified that such a class of
  functions is sufficiently general to generate (through
  transformations such as \eq{stress}) any pair of physically sensible
  holographic stress tensors at both boundaries.

\end{enumerate}

\begin{figure}[ht]
\centerline{\includegraphics[width=400pt, height=200pt]{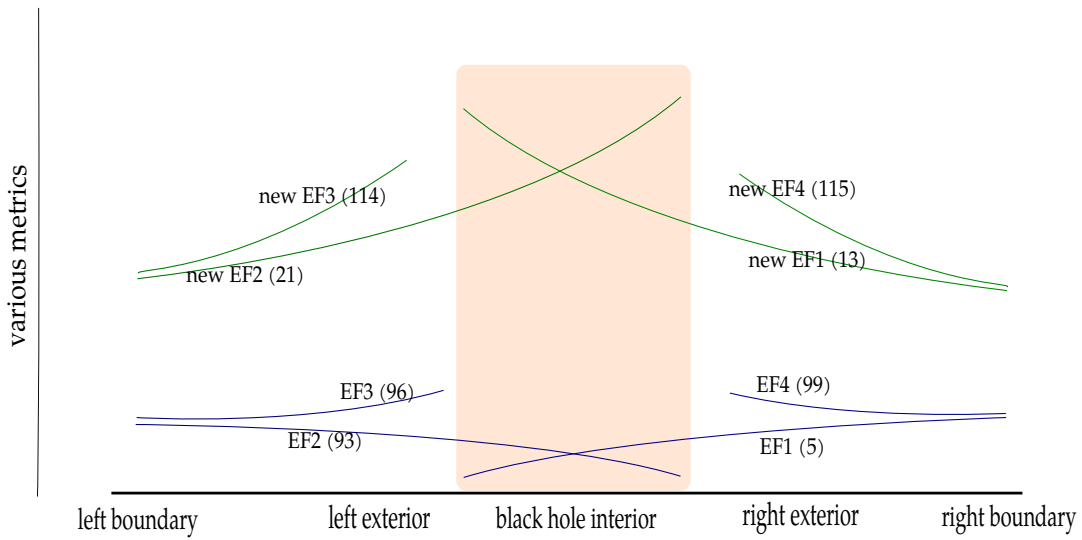}}
\caption{A schematic illustration of metrics in our paper related by
  trivial and nontrivial diffeomorphisms (see the definition
  \ref{def-nontrivial}).  The metrics \eq{EF}, \eq{EF2}, \eq{EF3} and
  \eq{EF4}, represented by the blue lines, define the eternal BTZ
  geometry; they are all related by trivial diffeomorphisms, which
  either do not extend to the boundaries or when they do, they become
  identity asymptotically. The metrics \eq{newmetric},
  \eq{newmetric-2}, \eq{newmetric-3} and \eq{newmetric-4}, represented
  by the green lines, define our new solution characterized by the
  functions $G_\pm, H_\pm$.  These are also all related by trivial
  diffeomorphisms, which satisfy the same criteria as above. The two
  sets however represent physically different metrics since they are
  related to each other by nontrivial diffeomorphisms; for instance,
  \eq{EF} and \eq{newmetric} are related by a diffeomorphism,
  schematically represented by their separation, which does not vanish
  (become identity) asymptotically.}
\label{fig-nontrivial}
\end{figure}

\subsubsection{\label{Dirac}Analogy with the Dirac monopole}

It is important to note that our new solutions can only be specified
in terms of a different metric in different coordinate charts which
are equivalent to each other.  This is analogous to case of the Dirac
monopole: the gauge field $A_\mu$ for a static U(1) magnetic monopole
of charge $q_m$ at the origin needs to be specified separately on two
separate coordinate charts:
\begin{align}
F= q_m \sin\theta\ d\theta\ d\phi: \;
A_N = q_m (1- \cos\theta)\ d\phi,\;
A_S = q_m (-1- \cos\theta)\ d\phi
\label{dirac}
\end{align}
Here ${\mathbb R}^3 - \{0\}$ is viewed as ${\mathbb R} \times S^2$
where $S^2$ is described by two coordinate charts $N_N$ and $N_S$
(such as obtained by a stereographic projection on to the plane) which
include all points of $S^2$ minus the south and north pole
respectively. $A_N^\theta$ vanishes (and is hence regular) at the
north pole $\theta=0$, but develops a string singularity at the south
pole $\theta=\pi$ (for each $r>0$).  Similarly, $A_S$ is regular at
the south pole, but has a string singularity at the north pole. The
important point to note is that in spite of appearances, $A_N$ and
$A_S$ describe the same gauge field in the region of overlap
$N_N \cap N_S$. This is because in this region, $A_N = A_S + d\chi$ where $\chi=2 q_m d\phi$
represents a pure gauge transformation for appropriately quantized
$q_m$ (Dirac quantization condition).

In the present case the metric \eq{newmetric} written in EF1, although
non-singular on the future horizon, is singular on the past horizon for
general $G_\pm$. In order to describe the metric in a neighbourhood of the past horizon, we must
switch to the metric in EF4. Similarly, in order to describe the
diffeomorphism at the bifurcation surface, we must use the metric
\eq{Kruskal} in the K5 coordinate chart.

\subsubsection{Summary of this subsection:} The metrics \eq{newmetric}, 
\eq{newmetric-2}, \eq{newmetric-3}, \eq{newmetric-4} and \eq{Kruskal},
valid in the coordinate charts EF1, EF2, EF3, EF4 and K5 respectively, define a
spacetime with a regular metric. The metrics are asymptotically
AdS$_3$ at both the right and left boundaries; the subleading terms in
the metric are determined by the solution generating diffeomorphisms
$G_\pm, H_\pm$ and can be chosen to fit boundary data specified by
arbitrary holographic stress tensors. A schematic representation
of our solution is presented in Figure \ref{fig-nontrivial}.

\subsection{\label{horizon}Horizon}

In Section \ref{SGD} we viewed the SGDs as a coordinate
transformation. Alternatively, however, we can also view the
diffeomorphism as an active movement of points: $x^M \to {\tilde x}^M$
$= x^M + \xi^M$. In this viewpoint, the future horizon $\l= \l_H=
\l_0$ (see \eq{lam-horizon}) on the right moves to
\begin{align}
\lt_H = G_+'(\vt)\ G_-'(\wt) \l_0,\;
\lt_{1,H} = H_+'(\ut)\  H_-'(\omt) \l_0
\label{warped-horizon} 
\end{align}

\begin{figure}[h]
\centerline{\includegraphics[width=180pt,height=150pt]{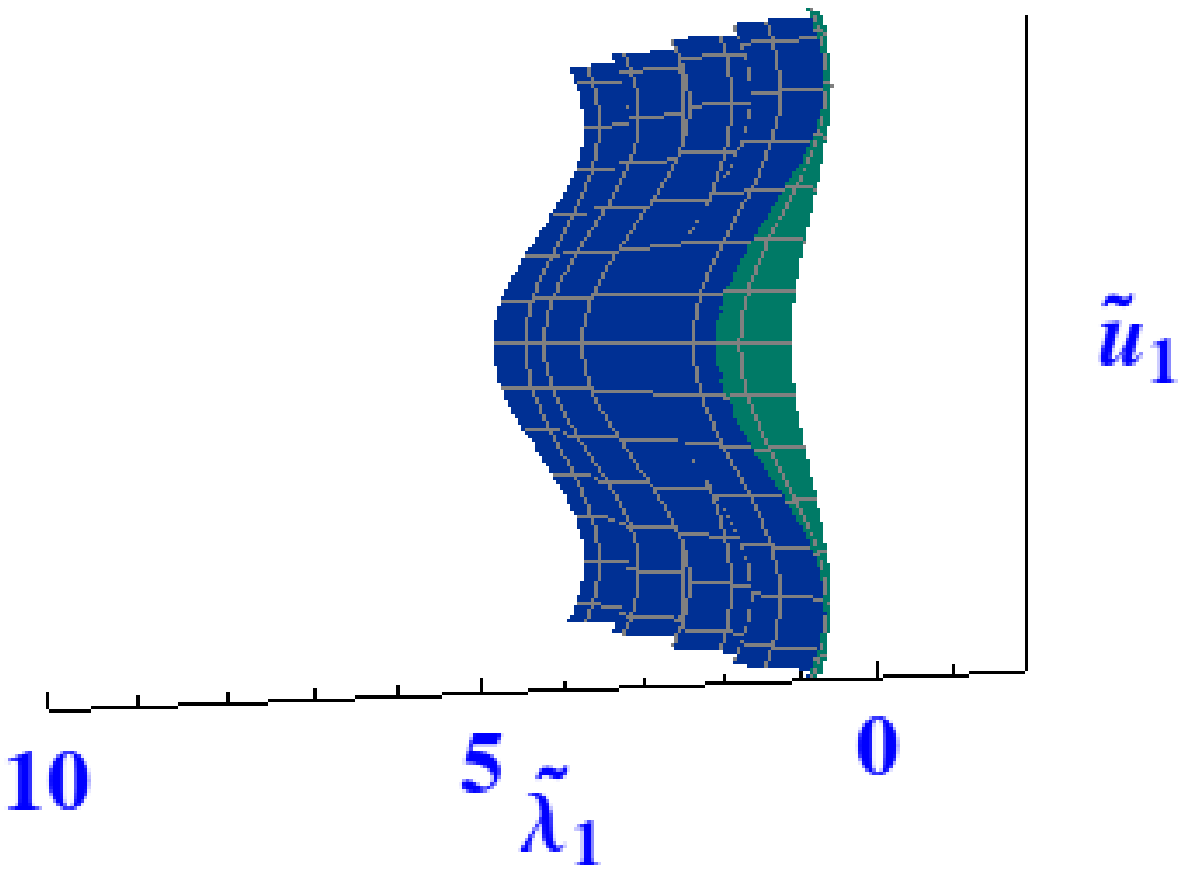}\kern15pt
\includegraphics[width=180pt,height=150pt]{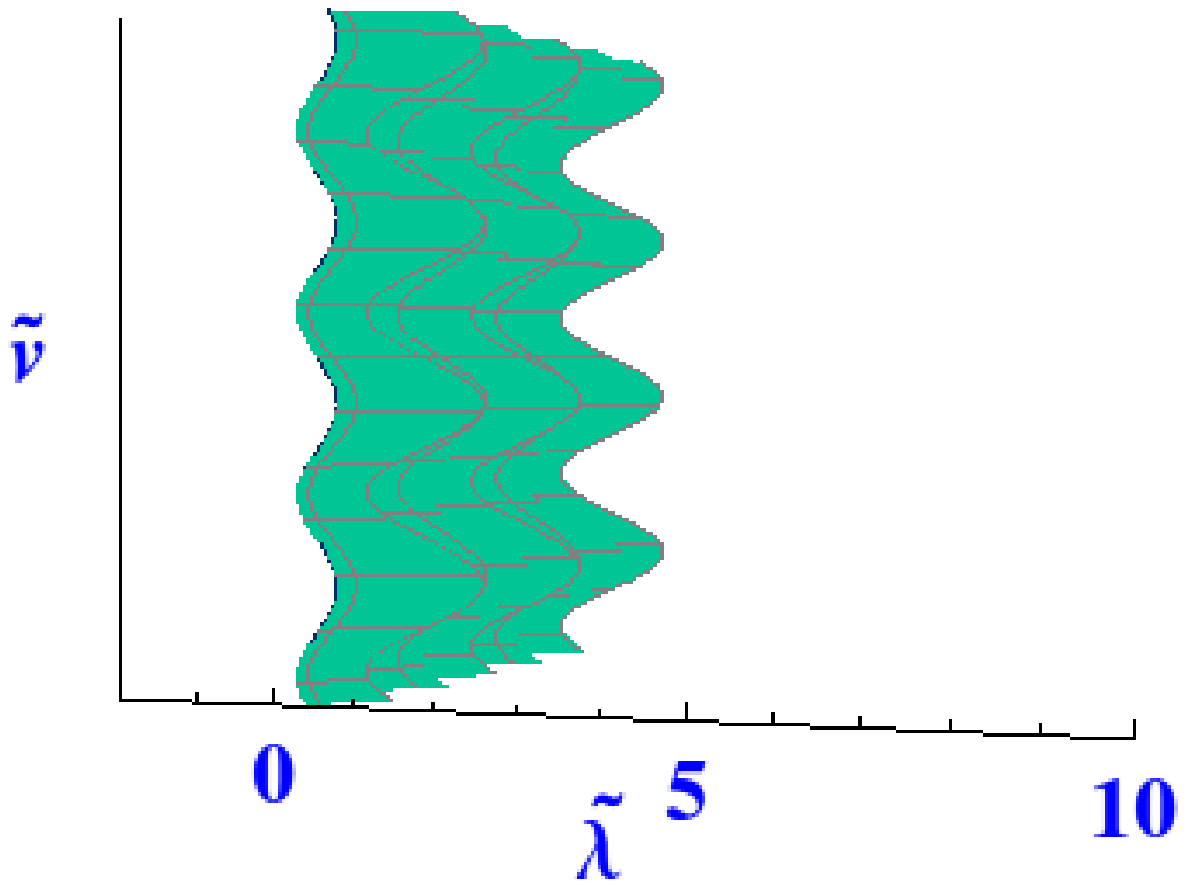}}
\caption{The figure on the right shows the location of the horizon on
  the right in the $\lt, \vt, \wt$ coordinates. The figure on the left
  shows the location of the horizon on the left in the $\tilde \l_1,
  \ut, \omt$ coordinates. These are described by \eq{warped-horizon}.
  These surfaces are diffeomorphic to the undeformed horizon
  \eq{lam-horizon} depicted in Figure \ref{fig-ef}. Although the horizon
has an undulating shape in our coordinate system, the 
expansion parameter, measured by the divergence of the area-form,  
vanishes (see Eq. \eq{no-divergence}).}
\label{fig-horizons}
\end{figure}

Similar statements can be made in the other coordinate charts. The
horizons represented this way are smooth but undulating (see figure
\ref{fig-horizons}).

The geometry of warped horizons in
\cite{Bhattacharyya:2008jc,Bhattacharyya:2008xc} was used to yield a
holographic prescription for computing local entropy current of a
fluid. In Section \ref{sec-S} we use a similar technology to compute a
holographic entropy in our case.
 
\subsection{On the nontriviality of solution generating
diffeomorphisms \label{nontrivial}}

It is natural to wonder how a metric such as \eq{newmetric} provides a
new solution since it is obtained by a diffeomorphism from \eq{EF};
however, the fact that the diffeomorphism \eq{diffeo} is
asymptotically nontrivial makes the new solution physically
distinct. Thus, in \eq{diffeo} $\lt$ remains different from $\l$ in
the asymptotic region.  Indeed, as we will see, the first subleading
term in the metric \eq{newmetric} carries nontrivial data about a
holographic stress tensor \eq{stress} on the right boundary.

Asymptotically AdS$_3$ diffeomorphisms were first discussed by Brown
and Henneaux \cite{Brown:1986nw} who showed that such transformation
led to an additional surface contribution to conserved charges of the
system. These observations were preceded by a general discussion of
such surface charges in the context of gauge theories and gravity in
\cite{Regge:1974zd,Wadia:1979yu,Gervais:1976ec}. These
authors identified asymptotically non-vanishing pure gauge transformations
as global charge rotations.

In the current AdS/CFT context, the surface charges are encapsulated
by the holographic stress tensors on the two boundaries. As we will
see shortly, they change nontrivially under the solution generating
diffeomorphisms (SGD's). In fact, the SGD's reduce to conformal
transformations on the boundary. As a result, the `global charge
rotations' mentioned above correspond to a {\it conformal}
transformation of the stress tensor. The important point is that
starting from a given constant stress tensor on each boundary,
the two independent SGD's can generate two {\it independent} and 
{\it completely general} stress tensors by this method.

We should note that the diffeomorphisms define a new theory in which
the appropriate choice of the IR cutoff surface is \eq{lamt-ir}. In
this description, the horizon becomes an undulating surface as in Fig
\ref{fig-horizons}. An equivalent (`active') viewpoint is to describe
the new geometry in terms of the old coordinates \eq{EF}, but to
change the IR-cutoff surface from \eq{lam-ir} to \eq{lamt-ir}. In
either case, the holographic stress tensor changes.

We conclude this section with the following definition of a nontrivial
diffeomorphism, which has been implicit in much of the above
discussion.

\subsubsection{\label{def-nontrivial}Definition} 

A local diffeomorphism which does not extend to either boundary (left
or right), or a diffeomorphism which extends to a boundary but
asymptotically approaches the identity diffeomorphism there, is called
a `trivial' diffeomorphism. Contrarily, a diffeomorphism which
extends to a boundary where it does not approach the identity
diffeomorphism, is called `nontrivial'. Quantitatively, a nontrivial
diffeomorphism ($f$) is one under which the holographic stress tensor
computed from the existing metric $g$ at the boundary is different
from that computed from the pulled back metric $f^*g$.

\section{The Dual Conformal Field Theory \label{CFT-dual}}

As we saw above, the SGD's reduce to conformal transformations at the
boundary. We will construct the CFT-dual to the new solutions
using the above idea.

Note that the eternal BTZ black hole geometry,
described by \eq{EF} and \eq{EF2}, corresponds to the following
thermofield double state
\cite{Maldacena:2001kr,Hartman:2013qma,Hubeny:2007xt,Caputa:2013eka}
\be
| \psi_0 \ran =  Z(\b_+, \b_-)^{-1/2} \sum_{n} \exp[-\b_+ E_{+,n}/2
- \b_- E_{-,n}/2] | n\ran  | n \ran
\label{thermofield-double}
\ee
The states $| n\ran  
\in {\cal H}$ denote all simultaneous eigenstates of
$H_\pm= (H \pm J)/2$ with eigenvalues $E_{\pm,n}$. 
$| \psi_0 \ran$ here is a pure state in $ {\cal H} \otimes {\cal
  H}$ obtained by the `purification' of the thermal state
\eq{thermal-state}. \footnote{For definiteness, we will sometimes
call the two Hilbert spaces ${\cal H}_L$ and ${\cal H}_R$, where
$L, R$ represent `left' and `right', corresponding to the two
exterior boundaries of the eternal BTZ. Indeed, $L, R$ also
have an alternative meaning. The left/right boundary of the eternal BTZ
geometry maps to the left/right Rindler wedge of the boundary of
 Poincare coordinates, respectively.\label{left-right}} 
 
\be
Z(\b_+, \b_-) = \Tr \rho_{\b_+, \b_-}\hspace{1cm}\mbox{with} \kern20pt
\rho_{\b_+, \b_-}= \exp[-\b_+ H_+ - \b_- H_-] = \exp[-\b(H 
+  \Omega  J)]
\label{thermal-state}
\ee
represents the grand canonical ensemble in ${\cal H}$ with inverse
temperature $\b$ and angular velocity $\Omega$ (which can be viewed as
the thermodynamic conjugate to the angular momentum $J$). Also $\b_\pm =
\b(1\pm \Omega)$. \footnote{The thermal state $\rho_{\b_+, \b_-}$ (see
\eq{thermal-state}) implies a field theory geometry where the light
cone directions have periods $\b_\pm$.}

Note that $| \psi_0 \ran$ is a pure state in $ {\cal H} \otimes {\cal
  H}$ , and is a `purification' of the thermal state
\eq{thermal-state}. 
\ni\underbar{The non-spinning BTZ:}
The CFT dual for the more familiar case of non-spinning eternal BTZ
black hole ($\Omega=0= J$) is the standard thermofield double:
\be
| \psi_{0,0} \ran =  Z(\b)^{-1/2} \sum_{n} \exp[-\b E_n/2]
| n\ran  | n \ran
\label{standard-thermofield-double}
\ee
where $| n \ran$ now denotes all eigenstates of $H$.
\footnote{An entanglement entropy for this state was calculated in
  \cite{Hartman:2013qma} and matched with a bulk geodesic calculation.
  This was generalized to the spinning eternal BTZ black hole in
  \cite{Caputa:2013eka}}

\paragraph{CFT duals of our solutions}
Following the arguments above
\eq{thermofield-double}, we claim 
that the CFT-duals to the new solutions described in
Section \ref{full-metric}  
are described by the following pure states in ${\cal H}
\otimes {\cal H}$:
\begin{align}
| \psi \ran = U_L U_R | \psi_0 \ran= 
 Z(\b_+, \b_-)^{-1/2}\sum_{n} \exp[-\b_+ E_{+,n}/2
- \b_- E_{-,n}/2]  U_L | n \ran U_R | n \ran
\label{thermofield-general}
\end{align} 
where $U_R$ is the unitary transformation which implements the
conformal transformations on the CFT on the right boundary
(characterized by
$G_\pm$), and $U_L$ is the unitary transformation which implements the
conformal transformations on the CFT on the left boundary (characterized by
$H_\pm$).  See Appendix \ref{unitary} for an explicit construction of a
unitary transformations $U_R$.

\gap1

\ni In the following sections, we will provide many checks for this proposal.
However, first we shall discuss how to compute various correlators
in the above state \eq{thermofield-general}.

\subsection{\label{corr-gen}Correlators}

Let us first consider correlators in the standard
thermofield double state \eq{thermofield-double}. 
It is known that correlators of one-sided CFT observables, 
say $O_R$, satisfy an AdS/CFT relation of the form \footnote{We
will mostly use unprimed labels, $P_1, P_2,...$ for points on
the spacetime of the `right' CFT, and primed labels, $P'_1, P'_2,...
$ for the space of the `left' CFT.}  
\be
\lan \psi_0 |   O_R(P_1) O_R(P_2) ... O_R(P_n) | \psi_0 \ran
\equiv \Tr\left( \rho_{\b_+, \b_-}  
O_R(P_1) O_R(P_2) ... O_R(P_n) \right)
=  G_{\rm bulk}(\bP_1, \bP_2,...\bP_n)
\label{RR-correlator}
\ee
where the bulk correlator $G_{bulk}$  
is computed from the (right exterior region of) 
a dual black hole geometry with 
temperature $T=1/\b$ and angular velocity $\Omega$. 
Two-sided correlators, similarly, satisfy a relation like 
\be
\lan \psi_0 |   O_R(P_1) O_R(P_2) ... O_R(P_m) 
O_L(P'_1)...O_L(P'_n) | \psi_0 \ran
=  G_{\rm bulk}(\bP_1, \bP_2,...\bP_m; \bP'_1,..., \bP'_n)
\label{RL-correlator}
\ee
where the bulk correlator on the RHS is computed from the two-sided geometry
of the eternal BTZ black hole \cite{Maldacena:2001kr,
Hartman:2013qma,Hubeny:2007xt,Caputa:2013eka}, 
represented in this paper by \eq{EF} and \eq{EF2}. The bold-faced label $\bP$ 
above represents an image of the field theory point $P$ on a
cut-off surface in the bulk under the usual 
AdS/CFT map. E.g. in the coordinates of \eq{EF}, the map is given by
\be
P \mapsto \bP\equiv (\l=\l_{ir}=1/\eps^2,P)
\label{bold-map}
\ee
where $\eps$ is the UV cut-off in the CFT, cf. \eq{lam-ir}). 
There is a similar map for the {\it left} boundary.\\
In particular, the holographic correspondence for
the two point functions of scalar operators 
can be written simply as \cite{Louko:2000tp}:
\begin{align}
\lan \psi_0 |   O_R(P) O_R(Q) | \psi_0 \ran
&= \Tr ( \rho_{\b_+, \b_-} O_R(P) O_R(Q)) =  \exp[-2h L(\bP,\bQ)]
\nonumber\\
\lan \psi_0 |   O_R(P) O_L(Q') | \psi_0 \ran
& =  \exp[-2h L(\bP,\bQ')]
\label{geodesics}
\end{align}
where $L(\bP,\bQ)$ is the length of the extremal geodesic
connecting $\bP$ and $\bQ$ (similarly with $L(\bP,\bQ')$).\\
It is easy to see that correlators in the new, transformed, state
$|\psi \ran$ \eq{thermofield-general} can be understood as correlators
of transformed operators in the old state $|\psi_0 \ran$, i.e.
\begin{align}
\lan \psi |   O_R(P_1)...O_R(P_m)O_L(P'_1)... O_L(P'_n) | \psi \ran
=  \lan\psi_0 |
\tilde O_R(P_1) ...\tilde O_R(P_m) \tilde O_L(P'_1) 
... \tilde O_L(P'_n) | \psi_0 \ran
\label{transform-corr}
\end{align}  
where 
\begin{align}
\tilde O_R(P) \equiv  U^\dagger_R O_R(P) U_R, \kern20pt
\tilde O_L(P') \equiv  U^\dagger_L O_L(P') U_L
\label{transform-op}
\end{align}
For a primary field $O_R$ with conformal dimensions $(h,\bar h)$, the
conformally transformed operator satisfies the relation
\be
\tilde O_R(\tilde v, \tilde w) 
= O_R(v,w) \left(\frac{dv}{d\vt}\right)^h 
\left(\frac{dw}{d\wt}\right)^{\bar h}
\label{primary}
\ee

\subsection{\label{checking}Strategy for checking AdS/CFT} 

To check the claim that the states \eq{thermofield-general} are
CFT-duals to the new bulk geometries found in Section \ref{full-metric}, 
we need to show a relation of the form
(cf. \eq{RL-correlator})
\begin{align}
\lan\psi_0 |
\tilde O_R(P_1) ...\tilde O_R(P_m) \tilde O_L(P'_1) 
... \tilde O_L(P'_n) | \psi_0 \ran
= \tilde G_{\rm bulk}(\tilde \bP_1, 
\tilde \bP_2,...\tilde \bP_m; \tilde \bP'_1,...,
\tilde \bP'_n)
\label{matching}
\end{align}  
where the RHS is computed in the new geometries. Here $\tilde \bP$
represents the image of the CFT point $P$, under AdS/CFT, on the
cut-off surface \eq{lamt-ir} in the new geometry. In the
language of \eq{newmetric}, the map is
\be
P \mapsto \tilde \bP= (\lt=\lt_{ir}=1/\eps^2,P)
\label{new-bold-map}
\ee
\ni\underbar{Two-point correlators:} In the 
particular case of two-point functions 
\begin{align}
\lan \psi_0 |   \tilde O_R(P) \tilde 
O_R(Q) | \psi_0 \ran
&= \Tr ( \rho_{\b_+, \b_-} \tilde O_R(P) \tilde O_R(Q)) =  
\exp[-2h \tilde L(\tilde \bP,\tilde \bQ)]
\nonumber\\
\lan \psi_0 |  \tilde  O_R(P) \tilde O_L(Q') | \psi_0 \ran
& =  \exp[-2h \tilde L(\tilde \bP, \tilde \bQ')]
\label{geodesics-new}
\end{align}
where $\tilde L(\tilde \bP,\tilde \bQ)$ is the length of the extremal
geodesic connecting $P$ and $Q$ in the new geometry (similarly with
$\tilde L(\tilde \bP, \tilde \bQ')$). The discerning reader may
justifiably wonder how a geodesic length in the new geometry can be
different from that in the original, eternal BTZ black hole geometry,
since the former is obtained by a diffeomorphism
from the latter; the point is that the bulk points $\tilde \bP$,
given by \eq{new-bold-map} are {\it not} the same as the bulk
points $\bP$ given by \eq{bold-map}. For example, a geodesic with endpoints
at a fixed IR cut-off $\lt= 1/\eps^2$ (both 
on the right exterior) corresponds, in the
eternal BTZ black hole, to a geodesic with two end-points 
at \eq{lamt-lam-ir} $\l= 1/(\eps^2 G_+'(\vt) G_-'(\wt))$.
As we will see below, it is this shift which ensures the equality
in \eq{geodesics-new}. This is one more instance of how our geometries
are nontrivially different from the original BTZ solution although
they are obtained by diffeomorphisms (see Section \ref{nontrivial}
for more detail). 

\section{Holographic Stress Tensor\label{sec-T}}

In this section we will discuss our first observable $O$: the stress
tensor. We will first consider the stress tensor of the boundary
theory on the right. The generalization to the stress tensor on the
left is trivial. The equation \eq{matching} now implies that we
should demand the following equality
\begin{align}
\lan\psi| T_{vv}(P) | \psi \ran
\equiv \Tr \left( \rho_{\b_+, \b_-} U_R^\dagger T_{vv}(P) U_R \right)
= \tilde T_{{\rm bulk},\vt\vt}(\tilde \bP)
\label{matching-stress}
\end{align}
and a similar equation for the right-moving stress tensor $T_{ww}(w)$.

\paragraph{Bulk} The RHS of this equation is simply the 
holographic stress tensor, computed in the new geometry
\eq{newmetric}. We use the definition of holographic stress tensor in
\cite{Balasubramanian:1999re,Skenderis:1999nb}:\footnote{We
drop the subscript {\tiny  bulk} from the bulk stress
tensor, as it should be obvious from the context whether we
are talking about the CFT stress tensor or the holographic
stress tensor.} 
\be
8\pi G_3 T_{\mu\nu}= \lim_{\eps\to 0}\left( 
K_{\mu\nu}- K h_{\mu\nu} -  h_{\mu\nu} \right)
\label{t-ren}
\ee 
where $h_{\mu\nu}$ is the induced metric on the cut-off surface
$\Sigma:\lt= \lt_{ir}=1/\eps^2$, chosen in accordance with 
\eq{new-bold-map} which is the natural one in the new
geometry (note that it is different from 
the cut-off surface implied by \eq{bold-map}). 
$K_{\mu\nu}$ and $K$ are respectively
the extrinsic curvature and its trace on $\Sigma$.
It is straightforward to do the explicit calculation; we find
that 
\begin{align}
8\pi G_3 T_{\vt\vt} &=\frac{L}{4} G_+'(\vt)^2 + \frac{3 G_+''(\vt)^2-2 G_+'(\vt) 
G_+'''(\vt)}{4G_+'(\vt)^2}, 
\nonumber\\
8\pi G_3 T_{\wt\wt} &=\frac{\bL}4 G_-'(\wt)^2 + \frac{3 G_-''(\wt)^2-2 G_-'(\wt) 
G_-'''(\wt)}{4G_-'(\wt)^2}
\label{stress}
\end{align}
This clearly looks like a conformal transformation of the
original stress tensor \eq{lam-stress}.  We will explicitly verify 
below that it agrees with the CFT calculation. The generalization
to $T_{ww}$ and to the stress tensors of the second CFT is
straightforward.This clearly has the form of a conformal transformation of the
original stress tensor \eq{lam-stress}.  We will explicitly verify
below in the CFT that it indeed is precisely a conformal
transformation, as demanded by \eq{matching-stress}. The
generalization of \eq{stress} to the stress tensors
$T_{\ut\ut}, T_{\omt\omt}$ of the second CFT
is straightforward.

In this paper, we will sometimes use the notation $T_R, \bar T_R$ for
$T_{\vt\vt}, T_{\wt,\wt}$, and $T_L, \bar T_L$ \footnote{$T_R, \bar
  T_R$ represent the left-moving and right-moving stress tensors on
  the Right CFT; similarly for $T_L, \bar T_L$.}  for $T_{\ut\ut},
T_{\omt\omt}$ respectively. It is clear that by appropriately choosing
the functions $G_\pm$ and $H_\pm$, any set of boundary stress tensors
$T_{R,L}, \bar T_{R,L}$ can be generated. This is how our solutions
described in Section \ref{full-metric} solve the boundary
value problem mentioned in the Introduction. 

\paragraph{CFT}

The unitary transformation in the LHS of \eq{matching-stress}, 
implements, by definition, the following conformal transformation
(see Appendix \ref{unitary} for more details) on the quantum
operator 
\be
U_R^\dagger T_{vv}(P) U_R 
= \bigg(\frac{\partial \vt}{\partial v}\bigg)^{-2}[T_{\vt\vt}(\vt)-
\frac{c}{12}S(v,\vt)]
\label{stress-transfmn}
\ee
From \eq{diffeo}, the relevant conformal transformation
here is $v= G_+(\vt)$. Using this, the definition
\eq{schwarzian} of the Schwarzian derivative $S(v, \vt)$, 
and the identification \cite{Brown:1986nw} 
\be
G_3= 3/(2c),
\label{newton}
\ee
we find that \eq{stress-transfmn} exactly agrees with \eq{stress}. 

\gap1

\ni This {\emph proves} the AdS-CFT equality \eq{matching-stress} for the
stress tensor.

\section{General two-point correlators\label{2-pt}}

In this section we will discuss general two-point correlators, both
from the bulk and CFT viewpoints following the steps outlined in
Section \ref{corr-gen}.

\subsection{Boundary-to-Boundary Geodesics}

As mentioned in \eq{geodesics}, the holographic calculation of a
two-point correlator reduces to computing the geodesic length between
the corresponding boundary points. We will first calculate correlators
in the thermofield double state \eq{thermofield-double}, which
involves computing geodesics in the eternal BTZ geometry \eq{EF}.

\subsubsection*{In the eternal BTZ geometry}

\ni\underbar{RL geodesic:} Let us consider a geodesic running from a
point $\bold{P}(1/\eps^2_R,v,w)$ on the right boundary to a 
point $\bold{Q'}=(1/\eps^2_L,u,\omega)$ on the left boundary.\footnote{For the
  calculation at hand we need to put $\eps_L = \eps_R= \eps$; however,
  we keep the two cutoffs independent for later convenience.}  As
shown in Section \ref{sec-poincare} (see \cite{Hartman:2013qma}) both the
right exterior ($\subset$ EF1) and the left exterior ($\subset$ EF2)
can be mapped to a single coordinate chart in Poincare coordinates.
Let the Poincare coordinates for ${\bf P}$ and ${\bf Q'}$, be
$(X_{+R},X_{-R},\zeta_R)$ and $(X_{+L},X_{-L},\zeta_L)$ respectively.
By using the coordinate transformations given in
(\ref{poincare-to-ef1}) and (\ref{poincare-to-ef2}), we find, upto the
first subleading order in $\eps_R$ and $\eps_L$, 
 \begin{eqnarray}
  \label{bp}
  &&X_{+R}=e^{\sqrt{L}v},\quad X_{-R}=-e^{-\sqrt{L}w}+L\eps_R^2 e^{-\sqrt{L}w},\quad\zeta_R^2=L\eps_R^2\,e^{\sqrt{L}(v-w)}\\
  &&X_{+L}=-e^{\sqrt{L}u}+L\eps_L^2 e^{\sqrt{L}u},\quad X_{-L}=e^{-\sqrt{L}\omega},\quad\zeta_L^2=L\eps_L^2\,e^{\sqrt{L}(u-\omega)}\nonumber\
 \end{eqnarray}
with $L=\bar{L}$.\footnote{For simplicity, we present the calculation
  here for $L=\bL$; the generalization to the spinning BTZ is
  straightforward.} The geodesic in Poincare coordinates is given by
\begin{eqnarray}
 X_+=A\tanh\tau+C,\quad X_-=B\tanh\tau+D,\quad \zeta=\frac{\sqrt{-AB}}{\cosh\tau}\nonumber\
\end{eqnarray}
where $\tau$ is the affine parameter, which takes the
values $\tau_R$ and  $\tau_L$ at ${\bf P}$ and ${\bf Q'}$ respectively. The
constants $A,B,C,D,\tau_L$ and $\tau_R$ are fixed by the endpoint coordinates
given above. 
In the limit $\eps_R,\eps_L\rightarrow0$, we obtain
\begin{eqnarray}
 \tau_{R}&=&\log\Big{[}\frac{e^{-(\sqrt{L}v+\sqrt{L}\omega)/2}}{\sqrt{2}}\sqrt{\frac{(e^{\sqrt{L}v}+e^{\sqrt{L}u})(e^{\sqrt{L}w}+e^{\sqrt{L}\omega})}{\lambda_0\eps_R^2}}\Big{]}\nonumber\\
 \tau_{L}&=&-\log\Big{[}\frac{e^{-\sqrt{L}(u+w)/2}}{\sqrt{2}}\sqrt{\frac{(e^{\sqrt{L}v}+e^{\sqrt{L}u})(e^{\sqrt{L}w}+e^{\sqrt{L}\omega})}{\lambda_0\eps_L^2}}\Big{]}\nonumber\
\end{eqnarray}
where $\l_0=L/2$ (see \eq{lam-horizon}). The geodesic length is now simply
given by the affine parameter length
\begin{eqnarray}
 L(\bold{P},\bold{Q'})=\tau_R-\tau_L=\log\left[\frac{4\cosh[\sqrt{L}(v-u)/2]\cosh[\sqrt{L}(w-\omega)/2]}{L\eps_R\eps_L}\right]\
 \label{geodesic}
\end{eqnarray}
For comparison with CFT correlators in the thermofield double, we will
put, in the above expression, $\eps_L= \eps_R =\eps$, where $\eps$ is
the (real space) UV cut-off in the CFT. 

\ni\underbar{RR geodesic:}
If we take the two boundary points on the same 
exterior region, say on the right, $\bold{P_1}(1/\eps_1^2,v_1,w_1)$ and
$\bold{P_2}(1/\eps_2^2,v_2,w_2)$, then the corresponding Poincare
coordinates are (using (\ref{poincare-to-ef1}))
\begin{eqnarray}
 \label{bp12}
  X_{+1}=e^{\sqrt{L}v_1}, &&\quad X_{-1}=-e^{-\sqrt{L}w_1}+L\eps_1^2 e^{-\sqrt{L}w_1},\quad\zeta_1^2=L\eps_1^2\,e^{\sqrt{L}(v_1-w_1)}\\
  X_{+2}=e^{\sqrt{L}v_2}, &&\quad X_{-2}=-e^{-\sqrt{L}w_2}+L\eps_2^2 e^{-\sqrt{L}w_2},\quad\zeta_2^2=L\eps_2^2\,e^{\sqrt{L}(v_2-w_2)}\nonumber\
 \end{eqnarray}
Following steps similar to above, we have, in the $\eps_1,\eps_2\rightarrow0$ limit,
\begin{eqnarray}
 \tau_{1}&=&\log\Big{[}\frac{e^{-(v_1+w_2)/2}}{\sqrt{2}}\sqrt{\frac{(e^{v_1}-e^{v_2})(-e^{w_1}+e^{w_2})}{\lambda_0\eps_1^2}}\Big{]}\nonumber\\
 \tau_{2}&=&-\log\Big{[}\frac{e^{-(v_1+w_1)/2}}{\sqrt{2}}\sqrt{\frac{(-e^{v_1}+e^{v_2})(e^{w_1}-e^{w_2})}{\lambda_0\eps_2^2}}\Big{]}\nonumber\
\end{eqnarray}
The geodesic length is then
\begin{eqnarray}
L(\bold{P_1},\bold{P_2})=\tau_{+1}-\tau_{+2}=\log\left[\frac{4\,\sinh[(v_1-v_2)/2]\sinh[(w_1-w_2)/2]}{L\epsilon_1\eps_2}\right]\
\label{geodesic12}
\end{eqnarray}
For comparison with CFT, we will put $\eps_1= \eps_2= \eps$.

\subsubsection*{In the new geometries}

As explained in Section \ref{behind}, the IR boundary in the new solutions, obtained by
the SGDs, is given by the equation \eq{lamt-ir} or
equivalently by \eq{lamt-lam-ir}, and analogous equations on the left.
This is encapsulated by the CFT-to-bulk map \eq{new-bold-map}. In 
case of the {\it RL geodesic}, the CFT endpoints $(P, Q')$ 
now translate to new boundary points $({\bf{\tilde P},{\tilde Q'}})$  
with the following new values of the old $(\l$, $\l_1)$ coordinates:  
\begin{equation}
\lambda \equiv
\frac{1}{\eps_R^2} =  \frac{1}{\epsilon^2 G'_+(\tilde{v}) G'_-(\tilde{w})},\; \quad \lambda_1 \equiv \frac{1}{\eps_L^2} =  \frac{1}{\epsilon^2 H'_+(\tilde{u}) H'_-(\tilde{\omega})}\;
\end{equation}
which just has the effect of conformally transforming  the 
boundary coordinates// $\epsilon_R=\eps \to 
\eps_R=\epsilon \sqrt{ G'_+(\tilde{v})G'_-(\tilde{w})}$, $\eps_L=
\eps \to \eps_L= \eps \sqrt{H'_+(\tilde{u}) H'_-(\tilde{\omega})}$. Using
these new values of $\eps_{L,R}$, we get
\begin{eqnarray}
 L(\bold{\tilde P},\bold{\tilde Q'})&=&\log\left[\frac{4\cosh[\sqrt{L}(G_+(\tilde{v})-H_+(\tilde{u}))/2]}{\sqrt{L}\eps\sqrt{G'_+(\tilde{v})H'_+(\tilde{u})}}\frac{\cosh[\sqrt{L}(G_-(\tilde{w})-H_-(\tilde{\omega}))/2]}{\sqrt{L}\eps\sqrt{ G'_-(\tilde{w})H'_-(\tilde{\omega})}}\right]\
 \label{geodesic1}
\end{eqnarray}
Similarly,
\begin{eqnarray}
L(\bold{\tilde P_1},\bold{\tilde P_2})
&=&\log\left[\frac{4\sinh[\sqrt{L}(G_{+}(\tilde{v}_1)-G_{+}(\tilde{v}_2))/2]}{\sqrt{L}\eps\sqrt{G'_+(\tilde{v}_1)G'_{+}(\tilde{v}_2)}}\frac{\sinh[\sqrt{L}(G_{-}(\tilde{w}_1)-G_{-}(\tilde{w}_2))/2]}{\sqrt{L}\eps\sqrt{ G'_{-}(\tilde{w}_1)G'_{-}(\tilde{w}_2)}}\right]\nonumber\\
 \label{geodesic121}
\end{eqnarray}

\subsection{General two-point correlators from CFT}

\subsubsection*{In the thermofield double state}

\underbar{RL correlator:} For the eternal BTZ string, the coordinate
transformations from the EF to Poincare (see Appendix
\ref{sec-poincare}) reduce, at the boundary, to a conformal
transformation from the Rindler to Minkowski coordinates, so that the
boundary of the right (left) exterior maps to the right (left) 
Rindler wedge \cite{Hartman:2013qma}. It is expedient
to compute the CFT correlations first in the Minkowski plane, and then
conformally transform the result to Rindler coordinates. Using this
method of \cite{Hartman:2013qma}, we get the following result
\begin{eqnarray}
\label{RL-corr}
\lan \psi_0| O(X_{+R},X_{-R})\, O(X_{+L},X_{-L})
| \psi_0 \ran &=&\frac{(\sqrt{L}e^{\sqrt{L}v})^h(\sqrt{L}e^{-\sqrt{L}w})^{\bar{h}}(-\sqrt{L}e^{\sqrt{L}u})^h(-\sqrt{L}e^{-\sqrt{L}\omega})^{\bar{h}}}{(\frac{e^{\sqrt{L}v}+e^{\sqrt{L}u}}{\epsilon})^{2h}(\frac{-e^{-\sqrt{L}w}-e^{-\sqrt{L}\omega}}{\epsilon})^{2\bar{h}}}\nonumber\\
 &=&\Big{(}\frac{4\cosh\,[\sqrt{L}(v-u)/2]\,\cosh\,[\sqrt{L}(w-\omega)/2]}{L\epsilon^2}\Big{)}^{-2h}\nonumber\
\end{eqnarray}
where the operator $O$ is assumed to have dimensions $(h,\bar{h})$ and
we have used a real space field theory cut-off $\eps$. We have related
the temperature of the CFT to $L(=\bL)$ by the equation
$\sqrt{L}=2\pi/\b$. //
It is easy to see that this correlator satisfies the relation \eq{geodesics}
\begin{eqnarray}
 \label{2pgeoRL}
 \lan \psi_0| O(X_{+R},X_{-R})\, O(X_{+L},X_{-L})
| \psi_0 \ran =e^{-2h L(\bold{P},\bold{Q})}\
 \end{eqnarray}
where in the expression on the right hand side for the geodesic length
(\ref{geodesic}), we use $\eps_R=\eps_L=\eps$ as explained before.

\underbar{RR correlator:} By
following steps similar to the above, the two-point
correlator between the points (\ref{bp12}) is given by 
\begin{eqnarray}
\lan \psi_0|\mathcal{O}(X_{+1},X_{-1})\, \mathcal{O}(X_{+2},X_{-2})| \psi_0 \ran
&=&\frac{(\sqrt{L}e^{\sqrt{L}v_1})^h(\sqrt{L}e^{-\sqrt{L}w_1})^{\bar{h}}(\sqrt{L}e^{\sqrt{L}v_2})^h(\sqrt{L}e^{-\sqrt{L}w_2})^{\bar{h}}}{(\frac{(e^{\sqrt{L}v_1}-e^{\sqrt{L}v_2}}{\epsilon})^{2h}(\frac{-e^{-\sqrt{L}w_1}+e^{-\sqrt{L}w_2}}{\epsilon})^{2\bar{h}}}\nonumber\\
 &=&\Big{(}\frac{4\sinh\,[\sqrt{L}(v_1-v_2)/2]\,\sinh\,[\sqrt{L}(w_1-w_2)/2]}{L\epsilon^2}\Big{)}^{-2h}\nonumber\
\end{eqnarray}
It follows, therefore, that 
 \begin{eqnarray}
 \label{2pgeo12}
  \lan \psi_0|\mathcal{O}(X_{+1},X_{-1})\, \mathcal{O}(X_{+2},X_{-2})
 | \psi_0 \ran&=&e^{-2hL(\bold{P_1},\bold{P_2})}\
 \end{eqnarray}
where, again, the geodesic length on the right hand side is read off
from \eq{geodesic1} with $\eps_1=\eps_2=\eps$.

 \subsubsection*{In the new states}

As explained in \eq{transform-corr}, correlators in the state $| \psi
\ran$ \eq{thermofield-general} can be computed by using a conformal
transformation \eq{primary} of the operators. The new correlator is,
therefore, found from the old one \eq{RL-corr} by a conformal
transformation of the boundary coordinates and an inclusion of the
Jacobian factors. The latter has, in fact, the effect of the
replacement $\epsilon^2 \to \epsilon^2 \sqrt{ G'_+(\tilde{v})
  G'_-(\tilde{w}) H'_+(\tilde{u})
  H'_-(\tilde{\omega})}$. With these ingredients, it is
straightforward to verify that \eq{geodesics-new} is satisfied.
Similar arguments apply to {\it RR} and {\it LL} correlators.

\section{Entanglement entropy\label{entang}}

We define an entangling region $A= A_R\cup A_L$, where $A_R$ is a half
line $(v-w)/2>x_R$ on the right boundary at `time' $(v+w)/2=t_R$ and
$A_L$ is a half line $(u-\omega)/2>x_L$ of the left boundary at `time'
$(u+\omega)/2=t_L$.  The boundary of the region $A$ consists of a
point $P(v_{\partial A},w_{\partial A})$ on the right and a point
$Q'(u_{\partial A},\omega_{\partial A})$ on the left, with coordinates
\begin{eqnarray}
\label{bpent}
P:\kern10pt v_{\partial A}&=&t_R+x_R,\quad w_{\partial A}=t_R-x_R\\
Q':\kern10pt 
u_{\partial A}&=&t_L+x_L,\quad \omega_{\partial A}=t_L-x_L\nonumber\
\end{eqnarray}

\subsection*{Bulk calculations}

\subsubsection*{In the BTZ geometry}

We calculate the entanglement entropy $S_A$ of the region A
using the holographic entanglement formula of \cite{Ryu:2006bv,Hubeny:2007xt}.
The HEE is given in terms of the geodesic length $L({\bf P,
Q'})$. The geodesic length, as calculated in (\ref{geodesic}), is 
\begin{eqnarray}
 L(\bold{P},\bold{Q'})=\log\left[\frac{4\cosh[\sqrt{L}(v_{\partial A}-u_{\partial A})/2]\cosh[\sqrt{L}(w_{\partial A}-\omega_{\partial A})/2]}{M\epsilon^2}\right]\nonumber\\
\end{eqnarray}
The HEE is then given by $S_A=L(\bold{P},\bold{Q'})/4G_3$. Using \eq{newton},
we get
\begin{eqnarray}
 S_A=\frac{c}{6}\log\left[\frac{4\cosh[\sqrt{L}((t_R+x_R)-(t_L+x_L))/2]\cosh[\sqrt{L}((t_R-x_R)-(t_L-x_L))/2]}{M\epsilon^2}\right]\
 \label{holoEE}
\end{eqnarray}
Note that for $x_R=x_L=0$ and $t=t_R=-t_L$ (which correspond to a
non-trivial time evolution in the geometry) the HEE \eq{holoEE}
reduces to
\begin{eqnarray}
S_{A}= \frac{c}{3}\log\left[\cosh\,\frac{2\pi t}{\b}\Big{]}+\frac{c}{3}\log\Big{[}\frac{\b/\pi}{\eps}\right]\
\label{HMresult}
\end{eqnarray}
which reproduces the result for the HEE in
\cite{Hartman:2013qma}.\footnote{The UV cutoff in
  \cite{Hartman:2013qma} is half of the cutoff, $\eps$ used here.}

\subsubsection*{In the new geometries}

The HEE corresponding to the conformally transformed state 
\eq{thermofield-general} is given by the length $L({\bf \tilde P, 
\tilde Q'})$ connecting the end-points $P$ and $Q'$ in the
new geometries described in Section \ref{full-metric}. Working on
lines similar to the derivation of \eq{geodesic12}, the HEE
is given by 
\begin{eqnarray}
 S_A&=&\frac{c}{6}\log\Big{[}\frac{4\cosh[\sqrt{L}(G_+(\tilde{t}_R+\tilde{x}_R)-H_+(\tilde{t}_L+\tilde{x}_L))/2]}{\sqrt{L}\eps\sqrt{G'_+(\tilde{t}_R+\tilde{x}_R)H'_+(\tilde{t}_L+\tilde{x}_L)}}\nonumber\\
 &&\quad\quad\quad\quad\frac{\cosh[\sqrt{L}(G_-(\tilde{t}_R-\tilde{x}_R)-H_-(\tilde{t}_L-\tilde{x}_L))/2]}{\sqrt{L}\eps\sqrt{ G'_-(\tilde{t}_R-\tilde{x}_R)H'_-(\tilde{t}_L-\tilde{x}_L)}}\Big{]}\
 \label{holoEE1}
\end{eqnarray}

\subsection*{CFT calculations}

\subsubsection*{In the thermofield double state}

The technique of calculating the entanglement entropy in the
thermofield double state is well-known \cite{Cardy:1986ie}.  The Renyi entanglement
entropy $S_A^{(n)}$ of the region A (\ref{bpent}) is given by the trace
of the $n^{th}$ power of the reduced density matrix $\rho^n_A$.  The
latter can be shown to be a Euclidean path integral on an $n$-sheeted Riemann
cylinder. This can then be calculated in terms of the two point correlator, on a complex
plane, of certain twist fields $\mathcal{O}$, with conformal dimensions 
\begin{eqnarray}
 h=\frac{c}{24}(n-1/n),\quad \bar{h}=\frac{c}{24}(n-1/n)\nonumber\\
\end{eqnarray}
inserted at the end-points $(P, Q')$ of A. The two-point correlator is
given by a calculation similar to that in the previous section. Thus,
\begin{eqnarray}
S_A^{(n)} &=&
\lan 
\mathcal{O}_R(v_{\partial A},w_{\partial A})\,\mathcal{O}_L(u_{\partial A},
\omega_{\partial A}) \ran \nonumber\\
 &=&\frac{(\sqrt{L})^{2h+2\bar{h}}}{(4\cosh[\sqrt{L}((t_R+x_R)-(t_L+x_L))/2]/\epsilon)^{2h}(\cosh[\sqrt{L}((t_R-x_R)-(t_L-x_L))/2]/\epsilon)^{2\bar{h}}}\nonumber\
\end{eqnarray}
The entanglement entropy $S_A=-\partial_n S_A^{(n)}|_{n=1}$ is
\begin{eqnarray}
 S_A=\frac{c}{6}\log\left[\frac{4\cosh[\sqrt{L}((t_R+x_R)-(t_L+x_L))/2]\cosh[\sqrt{L}((t_R-x_R)-(t_L-x_L))/2]}{L\epsilon^2}\right]\
 \label{CFTEE}
\end{eqnarray}
This proves that the CFT entanglement entropy and holographic entanglement
entropy(\ref{holoEE}) are equal.

\subsubsection*{In the new states}

The EE of the region A, computed in the new state  
\eq{thermofield-general}, is given in terms of the 
conformally transformed two-point function described in 
\eq{transform-corr}. The conformally transformed points are given by
\begin{eqnarray}
 v_{\partial A}=G_+(\tilde{v}_{\partial A})=G_+(\tilde{t}_R+\tilde{x}_R), &&\quad w=G_-(\tilde{w}_{\partial A})=G_-(\tilde{t}_R-\tilde{x}_R)\nonumber\\
 u_{\partial A}=H_+(\tilde{u}_{\partial A})=H_+(\tilde{t}_L+\tilde{x}_L), &&\quad \omega=H_-(\tilde{\omega}_{\partial A})=H_-(\tilde{t}_L-\tilde{x}_L)\nonumber\
\end{eqnarray}
It follows that the entanglement entropy is
\begin{eqnarray}
 S_{A,CFT}&=&\frac{c}{6}\log\Big{[}\frac{4\cosh[\sqrt{L}(G_+(\tilde{t}_R+\tilde{x}_R)-H_+(\tilde{t}_L+\tilde{x}_L))/2]}{\eps\sqrt{L}\sqrt{G'_+(\tilde{t}_R+\tilde{x}_R)H'_+(\tilde{t}_L+\tilde{x}_L)}}\nonumber\\
 &&\quad\quad\quad\quad\frac{\cosh[\sqrt{L}(G_-(\tilde{t}_R-\tilde{x}_R)-H_-(\tilde{t}_L-\tilde{x}_L))/2]}{\eps\sqrt{L}\sqrt{ G'_-(\tilde{t}_R-\tilde{x}_R)H'_-(\tilde{t}_L-\tilde{x}_L)}}\Big{]}\
 \label{CFTEE1}
\end{eqnarray}
which matches with the HEE (\ref{holoEE1}).

\subsection{\label{example}Dynamical entanglement entropy
in a specific new geometry}

We now compute the entanglement entropy in an illustrative geometry
specified by a particular choice of the functions $G_\pm$ and $H_\pm$.
In this example, we take
\begin{eqnarray}
x_R=0,\quad t_R=t,\quad x_L=0,\quad t_L=-t\nonumber\ 
\end{eqnarray}
For simplicity, we consider $G_\pm$ and $H_\pm$ which satisfy
\begin{eqnarray}
 G_+(x)\equiv G_-(x)\equiv G(x), \quad H_+(x)\equiv H_-(x)\equiv H(x)\nonumber\
\end{eqnarray}
With the transformations given above, we have
\begin{eqnarray}
\tilde{x}_R=0,\quad\tilde{v}_{\partial A}=\tilde{w}_{\partial A}=\tilde{t}_R=\tilde{t},\quad\quad \tilde{x}_L=0,\quad\tilde{u}_{\partial A}=\tilde{\omega}_{\partial A}=\tilde{t}_L=-\tilde{t}\
\label{ttilde}
\end{eqnarray}
The expression for the HEE  (\ref{holoEE1}) then reduces to
\begin{eqnarray}
S_A=\frac{c}{3}\log\left[\frac{2\cosh[\sqrt{L}(G(\tilde{t})+H_1(\tilde{t}))/2]}{\eps\sqrt{L}\sqrt{G'(\tilde{t})H'_1(\tilde{t})}}\right]\
\label{holoEEHMG}
\end{eqnarray}
where we have defined the notation $-H(-\tilde{t})=H_1(\tilde{t})$. 

\begin{figure}[H]
\centerline{\includegraphics[scale=.2]{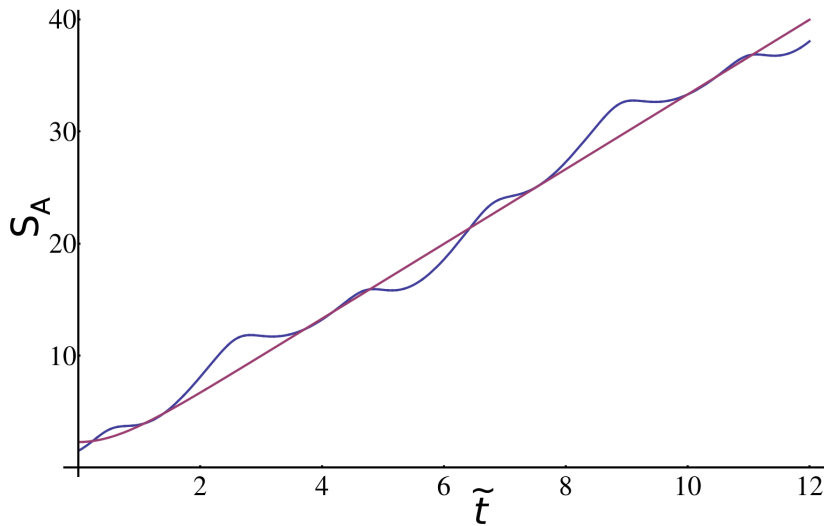}}
\caption{Time evolution of HEE. The red-line represents the linear growth of HEE for a region consisting of spatial half-lines of both sides of a constant 2-sided BTZ geometry.
The blue-line represents the HEE growth of the region consisting of half-lines of both sides of the SGD transformed geometry, for $G(\tilde{t})=\tilde{t}+\frac{1}{6}\cos(3\tilde{t})$ and $H_1(\tilde{t})=\tilde{t}+\frac{3}{5}\sin(\tilde{t})$. The undulating curve can be explained in terms of the quasiparticle
picture 
of \cite{Calabrese:2005in}; the entanglement
entropy departs from its usual linear behaviour as the
quasiparticle pairs locally go out and back in to the entangling region 
as the region is subjected
to a conformal transformation.}
\label{HEEplot}
\end{figure}

\section{Entropy\label{sec-S}} 

As discussed in previous sections, our solutions of Section
\ref{full-metric} are characterized by a smooth, albeit undulating,
horizon (see Figure \ref{fig-horizons}).  This allows us, following
\cite{Bhattacharyya:2008xc}, to define a holographic entropy
current. We will first review the equilibrium situation (static black
string), and then describe the calculation for the general,
time-dependent solution. We will include a comparison with CFT
calculations in both cases.

\subsection{Equilibrium}

\paragraph{Bulk calculation:}
In case $L= \bL=$ constant, our solutions represent BTZ black strings
\eq{EF} with a horizon  at $\l=\l_0$. The horizon ${\cal H}$ 
is a two-dimensional null surface, described by the metric
\be
ds^2|_{{\cal H}} \equiv H_{\mu\nu}dx^{\mu}dx^{\nu}
=  \left(\sqrt{L}dv/2 - \sqrt{\bL} dw/2\right)^2
\label{induced}
\ee
Since the normal to ${\cal H}$ at any point, given by 
$n^M = \del^M \l (M=\{\l, v,w\})$, also lies on ${\cal H}$,
${\cal H}$ possesses a natural coordinate system $(\tau, \a)$
where $\a$ labels  the one-parameter family of null geodesics,
and  $\tau$ measures the affine distance along the geodesics. 
In such a coordinate system, we get, by construction 
\be
ds^2|_{{\cal H}} = g d\alpha^2
\label{natural}
\ee
The area 1-form and the entropy current on the horizon are defined
by the equations \cite{Bhattacharyya:2008xc}
\footnote{Our  convention for $\eps_{\mu\nu}$ is $\epsilon_{vw}=-1$.},
\begin{align}
a \equiv 4 G_3\epsilon_{\mu\nu}J_S^{\mu}dx^{\nu} = \sqrt{g} d\alpha,
\label{area}
\end{align}
By inspection, from \eq{induced} and \eq{natural}, we find the
following expressions for the area-form and the entropy current
\begin{align}
a &=  \sqrt{L}dv/2 - \sqrt{\bL} dw/2
\nonumber \\ 
J_s^{v} &= \frac{1}{8G_3} \sqrt{\bL},\;
J_S^{w} = \frac{1}{8G_3} \sqrt L    
\label{area-val}
\end{align} 
The holographic entropy current on the boundary ${\cal B}$ is obtained
by using a map $f:{\cal B} \to {\cal H}$ and pulling back the
area-form (or alternatively the entropy current $J_{S,\mu}$) from
the horizon to the boundary. It turns out \footnote{The map $f$
is defined by shooting `radial' null geodesics inwards from the 
boundary, and is found to be of the form $f:(\l_{ir}, v, w)
\mapsto$ $(\l_{ir}, v+ C_1, w+C_2)$.}
that the natural pull back retains the form of the area-form
or entropy current, namely the expressions \eq{area-val}
still hold at the boundary.

To find the entropy density, we define the boundary coordinates $t=
(v+w)/2, x= (v-w)/2$ (see Section \ref{entang}), (so that
\eq{bdry-lam} has the canonical form $-dt^2 + dx^2$). With this the
entropy density becomes
\be
s \equiv J_S^T = \frac{1}{8G_3} \left( \sqrt L  + 
\sqrt{\bL} \right)
\label{cardy-b}
\ee

\paragraph{CFT calculation:}
 
The entropy density from the Cardy formula is \footnote{Recall that
  both $T_{vv}, T_{ww}$ are constant in this case. The more familiar
  form of \eq{cardy}, for a circular spatial direction of length
  $2\pi$, is obtained by putting $S= 2\pi s$, $L_0 = 2\pi T_{vv}$, and
  $\bar L_0= 2\pi T_{ww}$, which gives $S= 2\pi(\sqrt{c L_0/6}
+ \sqrt{c \bL_0/6}).$ } 
\be s= \sqrt{c \pi T_{vv}/3} + \sqrt{c \pi T_{ww}/3}
\label{cardy}
\ee
Using the identification \eq{newton} and \eq{lam-stress}, 
we can easily see that the two expressions \eq{cardy-b}
and \eq{cardy} exactly match. 

\subsection{New metrics: non-equilibrium entropy}

\paragraph{Bulk calculation:} We will now follow a similar procedure
as above, for the general solution in Section \ref{full-metric}. 
We find that (in coordinate chart EF1)
\be
ds^2|_\mathcal{H}=\frac14 d\alpha^2=\frac14 (\sqrt{L}G_+'(\vt)d\vt-\sqrt{\bL}G_-'(\wt)d\wt)^2
\ee
leading to the following area one form on the horizon
\be
a= \frac12 \sqrt{L}G_+'(\vt)d\vt-\frac12\sqrt{\bL}G_-'(\wt)d\wt
\label{areaform}
\ee
Note that this could alternatively be obtained from the area
form in \eq{area-val} by a diffeomorphism. 
The resulting expression for the entropy current, following the
steps above, is
\be
\Jt_s^{\vt} = \frac{1}{8G_3} \sqrt{\bL}  G'_-(\wt),\;
\Jt_S^{\wt} = \frac{1}{8G_3} \sqrt L  G'_+(\vt)   
\ee
Let us define, as before, the spacetime coordinates as $\tilde x,
\tilde t$ with $(\tilde v, \tilde w)$ = $\tilde t \pm \tilde x$.
The entropy density is then given by 
\be
\tilde s= \Jt_S^{\tilde t}=\frac{1}{4G_3 }\bigg(\frac12 
\sqrt L G_+'(\vt)+\frac12 \sqrt \bL G_-'(\wt)\bigg)
\label{current}
\ee
Note that the entropy current is  divergenceless
\be
\partial_\mu {\Jt}^{\mu}_S=
\partial_{\vt}{\Jt}^{\vt}_S+\partial_{\wt}\Jt^{\wt}_S=0
\label{no-divergence}
\ee
This has two implications:

\begin{enumerate}

\item \underbar{No dissipation:}
We have entropy transfers between different regions with no
net entropy loss or production (see Figure \ref{entropy}).

\begin{figure}[H]
\centerline{\includegraphics[scale=.5]{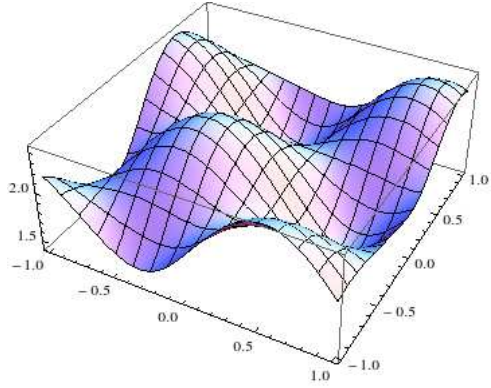}}
\caption{The undulating horizon of Figure \ref{fig-ef} leads to the
  non-trivial entropy current \eq{current}. In this figure, we plot the
  entropy density $\tilde s$ as a function of $\vt, \wt$ for the right
  CFT. Note that although the entropy density fluctuates, the
entropy flow here is such that there is no net entropy
production (or destruction) (see Eq. \eq{no-divergence}).}
\label{entropy}
\end{figure}

\item \underbar{Total entropy is not changed by  the
conformal transformation:}
The other implication is that the integrated entropy 
over a space-like (or null) slice $\Sigma$  
\be
\tilde S= \int_{\Sigma} \eps_{\mu\nu} J_S^\mu d\sigma^\nu
\ee
is independent of the choice of the slice. In particular,
choosing the slice to be $\Sigma_0:  t=v+w= 0$, we get
\begin{align}
\tilde S &= 
\frac1{8G_3}\ \int_{\Sigma_0} \left(\sqrt L G'_+(\vt) d\vt - \sqrt {\bL} G_-'(\wt) 
d\wt \right)
=\frac1{8G_3}\   \int_{\Sigma_0} \left(\sqrt L dv - \sqrt {\bL}  
dw \right) 
\\
\;& =\frac1{8G_3}\    \int dx \left(\sqrt L  + \sqrt {\bL}  \right) 
= \int dx\ s = S
\label{manipulate-S}
\end{align}
Hence although the entropy density is clearly transformed,
the total  entropy is not changed by the
conformal transformation.   

\end{enumerate}

\subsubsection*{CFT calculation:}

In a non-equilibrium situation, there is no natural notion of an
entropy. However under the adiabatic approximation, the instantaneous
eigenstates of a time-dependent Hamiltonian are a fair representation
of the actual time-dependent wave functions. The consequent energy
level density can thus be used to define an approximate time-dependent
entropy.  Generalizing this principle to slow time {\it and}
space variations, and applying this to the stress tensor, one
expects a space-time dependent version of \eq{cardy}, namely
\be
\tilde s=\sqrt{\frac{\pi c}{3}\tilde T_{\vt\vt}}+
\sqrt{\frac{\pi c}{3}\tilde T_{\wt\wt}}
\label{adiab}
\ee
where the stress tensors are given by \eq{stress}.
Since we have made the adiabatic approximation, we expect
the above formula to be valid only up to the leading order
of space and time derivatives. Under this approximation,
we have 
\begin{align}
8\pi G_3 T_{\vt\vt} &=\frac{L}{4} G_+'(\vt)^2,
\quad 
8\pi G_3 T_{\wt\wt} =\frac{\bL}4 G_-'(\wt)^2
\label{stress-adiab}
\end{align}
which exactly agrees with the holographic entropy density
in \eq{current}. \footnote{Note that 
throughout this paper, we have not used the adiabatic
approximation anywhere else. Thus, it is unsatisfactory
to use this approximation here. It is, in fact,
tempting to believe that the entropy density in \eq{current},
and not that in \eq{adiab}, actually gives the CFT entropy in general;
however, this requires more investigation.}  

\gap1

\underbar{Total entropy for ${\cal H}_R$ is unchanged
by the conformal transformation:}

\gap1

Under the conformal transformation \eq{transform-op}, the reduced
density matrix $\rho_R$ is changed by a unitary transformation: 
\be
\rho_R = \Tr_{{\cal H}_L} | \psi \ran \lan \psi | = U_R\, \rho_{0,R}\,
U_R^\dagger, \kern10pt  \rho_{0,R}= \Tr_{{\cal H}_L} | \psi_0 \ran \lan \psi_0 |
\label{rho-r}
\ee
The total entropy of the system after the transformation is given by
the von Neumann entropy $\tilde S= -\Tr \rho_R \ln \rho_R$ which,
therefore, is equal to the entropy before; it is unchanged by the
unitary transformation.

\section{Conclusion and open questions\label{discuss}}

In this paper we have solved the boundary value problem for 3D gravity
(with $\Lambda<0$) with independent boundary data on two
asymptotically AdS$_3$ exterior geometries. The boundary data,
specified in the form of arbitrary holographic stress tensors,
yields spacetimes with wormholes, {\it i.e.} with exterior regions
connected across smooth horizons. The explicit
metrics are constructed by the technique of solution generating
diffeomorphisms (SGD) from the eternal BTZ black string. By using the
fact that the SGD's reduce to conformal transformations at both
boundaries, we claim that the dual CFT states are specific
time-dependent entangled states which are conformal transformations of the
standard thermofield double. We compute various correlators and a
dynamical entanglement entropy, in the bulk and in the CFT, to provide
evidence for the duality. We also arrive at an expression for a
non-equilibrium entropy function from the area-form on the horizon of
these geometries.

Our work has implications for a number of other issues. We briefly
discuss two of them below; a detailed study of these is left to future
work.

\subsection{\label{EPR}ER=EPR}

As mentioned above, our work constructs an infinite family of AdS-CFT
dual pairs in which quantum states entangling two CFTs are
holographically dual to spacetimes containing a wormhole region which
connects the two exteriors.  Both the quantum states and the wormhole
geometries are explicitly constructed (see eqns.
\eq{thermofield-general} and (\ref{newmetric},\ref{newmetric-2})).
Our examples generalize the construction in 
 \cite{Maldacena:2001kr,Hartman:2013qma,Caputa:2013eka}\footnote{See \cite{Maldacena:2013xja,
Shenker:2013pqa,Shenker:2013yza}} 
(for other remarks on  unitary transformations of the
  thermofield double and related geometries see \cite{Shenker:2013pqa,Marolf:2013dba,
  Shenker:2013yza,Avery:2013bea,Balasubramanian:2014gla}) and provide
an infinite family of examples of the relation ER=EPR, proposed in
\cite{Maldacena:2013xja}. Since this relation has been extensively
discussed and debated in the literature (\cite{Shenker:2013pqa,Shenker:2013yza,
Avery:2013bea,Balasubramanian:2014gla}), we would like to make
some specific points pertaining to some of these discussions.

\subsubsection*{\label{no-suppression}RR correlators vs RL correlators}

It has been argued in \cite{Balasubramanian:2014gla},\cite{Shenker:2013pqa} and
\cite{Shenker:2013yza} that for typical
entangled states connecting two CFTs, $\cal{H}_R$ and $\cal{H}_L$,
correlators involving operators on the left and the right are
suppressed relative to those involving operators all on the right.  In
particular, according to \cite{Balasubramanian:2014gla}, correlators of the form $\lan O_R
O_L \ran$ are of the order $e^{-S} \lan O_R O_R \ran$, where $S$ is
the entropy of the right sided Hilbert space.

In Section \ref{2-pt} we have computed general two-point functions,
both of the kind $\lan O_R(P) O_R(Q) \ran$ and $\lan O_R(P) O_L(Q')
\ran$.\footnote{We use unprimed labels for operators on the right and
  primed labels for those on the left.} In case of the eternal BTZ
(dual to the standard thermofield double), an inspection of
\eq{geodesic} and \eq{geodesic12} suggests that as the boundary point
$\bP$ goes off to infinity, the $\cosh$ and $\sinh$ factors tend to be
equal, thus $L(\bP,\bQ) \approx L(\bP, \bQ')$, thus there is no extra
suppression in the two-sided correlator $\lan O_R\,O_L \ran$. Of
course, such a statement, regarding the standard thermofield double,
has been regarded as somewhat of a special nature.

We are therefore naturally led to ask: what happens in case of the new
solutions found in this paper? The geodesic lengths
$L(\bP, \bQ)$ and $L(\bP, \bQ')$ are now given by \eq{geodesic1} and
\eq{geodesic121}. Once again, if the point P goes off towards the
boundary of the Poincare plane, $\vt \to \infty$. Hence $G_+(\vt) \to
\infty$ (since $G_+$ is a monotonically increasing function). Hence,
both the geodesic lengths approach each other. Thus, we do not see any
peculiar additional suppression, even for our general entangled state,
arising when the second point of the correlation function is moved
from the right to the left CFT.

\subsubsection*{On the genericity of our family of examples} 

We start with the following Lemma.

\ni {\it Lemma:}
Any state $\in {\cal H} \otimes {\cal H}$,
\be 
| \Psi \ran = \sum_{i,j} C_{ij} | i \ran | j \ran, \kern5pt  C_{ij}\in 
{\mathbb C}, 
\label{entang-gen}
\ee
can be expressed in the form
\be 
| \Psi \ran = \sum_{i,j,n} e^{-\l_n} U_{L,in} U_{R,jn}  | n \ran | n \ran 
\label{entang-gen-a}
\ee  
where  $U_R, U_L$ are  two unitary operators and
$\l_n \ge 0$. 

\gap1

\ni {\it Proof:} Using the canonical map ${\cal H} \otimes {\cal H}$ $
\to {\cal H} \otimes {\cal H}^* $, we can regard the above state $|
\Psi \ran$ as an operator ${\bf \Psi}$ in ${\cal H}$, with matrix
elements $C_{ij}$. Using the singular value decomposition theorem on a
general complex matrix, we can write $C= U_L D U_R^\dagger$ where $D$
is a diagonal matrix with real, non-negative entries. By denoting $D$
as diag[$e^{-\l_n}$], we get \eq{entang-gen-a}.

\gap1

The state \eq{entang-gen-a} can be regarded as a thermofield double
with Hamiltonian $H= $$ \sum_n \l_n/\b$$ | n \ran \lan n |$
transformed by unitary operators $U_L$ on the left and by $U_R$ on the
right. Thus, the above Lemma suggests that the most general entangled
state \eq{entang-gen} can be written as a unitary transformation of
{\it some} thermofield double state. Now, note that the state
\eq{entang-gen-a} is of the same general form as that of
\eq{thermofield-general} discussed in this paper. {\it Are our states
  \eq{thermofield-general} the most general entangled states then?}

The answer is no, since the $U_{L,R}$ we use are made of Virasoro
generators (see Appendix \ref{unitary}), and are not the most general
unitaries of \eq{entang-gen-a}. However, in spite of this restriction,
it is clear that the states \eq{thermofield-general} do form a fairly
general class. Furthermore, if the states \eq{entang-gen-a} are states
in which only the stress tensor is excited, then indeed these states
are all contained in our class of states \eq{thermofield-general}. 
\footnote{If a CFT dual to pure gravity were to exist, then our states
  \eq{thermofield-general} in such a theory would indeed be the most
  general state of the form \eq{entang-gen}.  However, such a unitary
  theory is unlikely to exist \cite{Maloney:2007ud,Gaberdiel:2007ve},
  although chiral gravity theories which are dual to CFTs with only
  the Virasoro operator have been suggested (see,
  e.g. \cite{Afshar:2014rwa}).  We would like to thank Justin David
  for illuminating discussions on this point.}

\subsubsection*{Weakly entangled states}

To assess the genericity of our states, we ask a different question
now: do our set of states \eq{thermofield-general}, which are {\it
  all} explicitly dual to wormholes, include those with a very small
entanglement entropy $S$ for a given energy $E$?\footnote{This
  question was suggested to us by Sandip Trivedi.} The answer to this
question turns out to be yes. As we have noted in the remarks around
\eq{manipulate-S} and \eq{rho-r}, the entropy $S$, which is actually the
entanglement entropy of the right Hilbert space, is the same for all
our states. However, the same manipulations as in \eq{manipulate-S}
shows that the energy of these states are {\it not} the same; indeed
by choosing the derivatives $G_\pm'$ to be large, we can make the
energy of the transformed state to be much larger than that of the
standard thermofield double. Stated in another way, for states of a
given energy, our set of states includes states with entanglement
entropy much less than that of the thermofield double. This is
consistent with the proposal of \cite{Maldacena:2013xja} that even a small
entanglement is described by a wormhole geometry. 

\subsection{Generalizations and open questions}

It would be interesting to rephrase the results in this paper in terms
of the $SL(2,R) \times SL(2,R)$ Chern-Simons formulation
\cite{Witten:1988hc} of three-dimensional gravity. By the arguments in
\cite{Witten:1988hc}, all diffeomorphisms (together with appropriate
local Lorentz rotations) can be understood as gauge transformations of
the Chern-Simons theory. The Chern-Simons formulation has been
extended to the gauge group $SL(N,R)$ $\times$ $SL(N,R)$ to describe
higher spin theories \cite{Banados:1998gg,Campoleoni:2010zq}. It would
be interesting to see whether the nontrivial gauge transformations in
our paper generalizes to these higher gauge groups, and hence to
higher spin theories. A possible application of our methods in this
case would be to compute HEE by the prescriptions in
\cite{Ammon:2013hba} and \cite{deBoer:2013vca} in the nontrivial
higher spin geometries\footnote{We thank Rajesh Gopakumar for a
  discussion on this issue.}. We hope to come back to this issue
shortly.

The solutions presented in this paper are generated by SGDs which can
be regarded as forming a group ($\widetilde{\rm Vir}$
$\times$ $\widetilde{\rm Vir}$)$_L$ $\times$($\widetilde{\rm Vir}$
$\times$ $\widetilde{\rm Vir}$)$_R$.  Here the first $\widetilde{\rm
  Vir}$ denotes a group of SGDs which is parametrized by the function
$G_+$, and so on. As we emphasized in \eq{rho-r}, the reduced density
matrix on the right $\rho_R$ undergoes a unitary transformation under
this group of transformations, leaving the entropy unaltered. The
family of pure states \eq{thermofield-general} considered in this
paper can, therefore, be considered as an infinite family of
purifications of the class of density matrices $\rho_R$; it would be
interesting to see if these can be regarded as `micro-states' which can
`explain' the entropy of $\rho_R$. We hope to return to this issue
shortly.

It would also be interesting to use our work to explicitly study
various types of holographic quantum quenches involving quantum states
entangling two CFTs.\footnote{For a single CFT, a similar computation
  was done in, e.g., \cite{Roberts:2012aq,Ugajin:2013xxa}.} It would
be of particular interest to study limiting cases of our solutions
which correspond to shock-wave geometries.

\subsection*{Acknowledgement}

We would like to thank Atish Dabholkar, Justin David, Avinash Dhar,
Rajesh Gopakumar, Juan Maldacena, Shiraz Minwalla, Suvrat Raju, Ashoke
Sen, Lenny Susskind, Sandip Trivedi, Tomonori Ugajin and Spenta Wadia
for many useful discussions during the course of this work. R.S. would
like to thank Sachin Jain, Nilay Kundu and V. Umesh for
discussions. We are grateful to Justin David, Avinash Dhar, Rajesh
Gopakumar, Juan Maldacena, Lenny Susskind and Spenta Wadia for
important feedbacks on the manuscript.

\section*{Appendix}

\appendix

\section{Coordinate systems for the eternal BTZ geometry \label{charts}}

As we explained in the Introduction, the metric \eq{banados} describes
only the region exterior to the black hole horizon
\eq{banados-horizon}. As is well-known, for constant ($L, \bL$),
\eq{banados} describes a standard BTZ black hole with mass $M$ and
angular momentum $J$ given by \be L= 8G_3 (M+J),\; \bL= 8G_3 (M-J)
\label{btz-mass-ang}
\ee
In this section we will describe various coordinate
systems for this case.  In particular, we will describe
the five coordinate charts of Figure \ref{fig-ef1234}
which cover our spacetime. 

\subsection{\label{EFs}Eddington-Finkelstein coordinates}

\paragraph{EF1 (Right Exterior + Black Hole Interior)}

For a black hole with constant mass and
angular momentum, it is straightforward to find a coordinate
transformation from the ($z,x_+,x_-$) coordinates to a set of
Eddington Finkelstein coordinates which we denote by EF1 ($\lambda,v,y$)
\be
x_+=v-\frac{1}{2\sqrt{L}}\log\bigg(\frac{\lambda-\lambda_0}{\lambda+\lambda_0}\bigg),\hspace{0.2cm}
x_-=y+\sqrt{\frac{L}{\bL}}v-\frac{1}{2\sqrt{\bL}}\log\bigg(\frac{\lambda^2-\lambda_0^2}{4\bL}\bigg)
\ee
\be
z=\sqrt{\frac{2}{\lambda_0^2}\bigg(\lambda-\sqrt{\lambda^2-\lambda_0^2}\bigg)}
\ee
Under these transformations, we obtain the following metric 
\begin{eqnarray}
 ds^2=-\frac{2}{\bL}\lambda_0(\lambda-\lambda_0)dv^2+\frac{1}{\sqrt{\bL}} dvd\lambda+\frac{\bL}{4}dy^2-(\lambda-\lambda_0)dv dy
\end{eqnarray}
The horizon \eq{banados-horizon} of the metric \eq{banados} is now
located at $\lambda_0=\sqrt{L\bL}/2$.  The metric is obviously smooth
and describes the black hole interior.\footnote{\label{inner} It
  develops a coordinate singularity at the inner horizon $\l= -\l_0$;
  we do not discuss interpolation beyond the inner horizon in this
  paper, although it can be easily done.  In any case, there are
  strong reasons to believe that generically, the inner horizon and
  the associated exotic feature of infinitely repeating universes are
  unstable against even infinitesimal perturbations.} To achieve a
symmetry between the boundary coordinates, we find it convenient to
make one further coordinate transformation from $y$ to $w$
\be
y=w-\sqrt{\frac{L}{\bL}}v+\frac{1}{\sqrt{\bL}}\log\bigg(\frac{\lambda+\lambda_0}{2\sqrt{\bL}}\bigg)
\ee
In these new coordinates ($\lambda,v,w$), the metric becomes
\begin{eqnarray}
 ds^2=\frac{d\lambda^2}{4(\lambda+\lambda_0)^2}+\frac{L}{4}dv^2+
\frac{\bL}{4}dw^2-\lambda dv dw
 +\frac{\sqrt{L}}{2(\lambda+\lambda_0)}dv d\lambda+\frac{\sqrt{\bL}}{2(\lambda+\lambda_0)}dw d\lambda,
\end{eqnarray}
which is clearly symmetric between the `boundary coordinates' $v$ and $w$. 


\paragraph{EF2 (Left Exterior + Black Hole Interior)}~\\

We can invent a second set of coordinate transformations starting from the metric in the ($z,x_+,x_-$) coordinates
which would describe the left exterior region of the black hole along with the interior. This transformation is the following
\\
\be
x_+=u+\frac{1}{2\sqrt{L}}\log\bigg(\frac{\lambda_1-\lambda_0}{\lambda_1+\lambda_0}\bigg),\hspace{0.2cm}
x_-=y_1+\sqrt{\frac{L}{\bL}}u+\frac{1}{2\sqrt{\bL}}\log\bigg(\frac{\lambda_1^2-\lambda_0^2}{4\bL}\bigg)
\ee
\be
z=\sqrt{\frac{2}{\lambda_0^2}\bigg(\lambda_1-\sqrt{\lambda_1^2-\lambda_0^2}\bigg)}
\ee
The Eddington-Finkelstein metric obtained via this transformation is

\begin{eqnarray}
 ds^2=-\frac{2}{\bL}\lambda_0(\lambda_1-\lambda_0)du^2-\frac{1}{\sqrt\bL}dud\lambda_1+\frac{\bL}{4}dy_1^2-(\lambda_1-\lambda_0)du dy_1
\end{eqnarray}
As before, we make a further coordinate transformation $y_1$ to $\omega$ 
\be
 y_1=\omega-\sqrt{\frac{L}{\bL}}u-\frac{1}{\sqrt{\bL}}\log\bigg(\frac{\lambda_1+\lambda_0}{2\sqrt{\bL}}\bigg)
\ee
to obtain the following metric in the ($\lambda_1,u,\omega$) coordinates 
\begin{eqnarray}
 ds^2=\frac{d\lambda_1^2}{4(\lambda_1+\lambda_0)^2}+\frac{L}{4}du^2+\frac{\bL}{4}d\omega^2-\lambda_1du d\omega
 -\frac{\sqrt{\bL}}{2(\lambda_1+\lambda_0)}d\omega d\lambda_1-\frac{\sqrt{L}}{2(\lambda_1+\lambda_0)}du d\lambda_1
\label{EF2}
\end{eqnarray}

\paragraph{EF3 (Left Exterior + White Hole Interior)}~\\

Starting from $(z,x_+,x_-)$ coordinates, we do the following transformations
\be
x_+=v_1-\frac1{2\sqrt{L}}\log\bigg(\frac{\l_1-\l_0}{\l_1+\l_0}\bigg),\hspace{0.2cm}
x_-=w_1-\frac1{2\sqrt{\bL}}\log\bigg(\frac{\l_1-\l_0}{\l_1+\l_0}\bigg)
\ee
\be
z=\sqrt{\frac2{\l_0^2}\bigg(\l_1-\sqrt{\l_1^2-\l_0^2}\bigg)}
\ee
The metric obtained is
\be
ds^2=\frac{d\lambda_1^2}{4(\lambda_1+\lambda_0)^2}+\frac{L}{4}dv_1^2+\frac{\bL}{4}dw_1^2-\lambda_1 dv_1 dw_1
+\frac{\sqrt{L}}{2(\lambda_1+\lambda_0)}dv_1 d\lambda_1+\frac{\sqrt{\bL}}{2(\lambda_1+\lambda_0)}dw_1 d\lambda_1
\label{EF3}
\ee
This metric covers the left exterior and the white hole interior.

\paragraph{EF4(Right Exterior + White Hole Interior)}~\\
\\
Starting from $(z,x_+,x_-)$ coordinates, we do the following transformations
\be
x_+=u_1+\frac1{2\sqrt{L}}\log\bigg(\frac{\l-\l_0}{\l+\l_0}\bigg),\hspace{0.2cm}
x_-=\om_1+\frac1{2\sqrt{\bL}}\log\bigg(\frac{\l-\l_0}{\l+\l_0}\bigg)
\ee
\be
z=\sqrt{\frac2{\l_0^2}\bigg(\l-\sqrt{\l^2-\l_0^2}\bigg)}\\
\ee
The metric obtained is
\be
ds^2=\frac{d\lambda^2}{4(\lambda+\lambda_0)^2}+\frac{L}{4}du_1^2+\frac{\bL}{4}d\om_1^2-\lambda du_1 d\om_1
-\frac{\sqrt{L}}{2(\lambda+\lambda_0)}du_1 d\lambda-\frac{\sqrt{\bL}}{2(\lambda+\lambda_0)}d\om_1 d\lambda
\label{EF4}
\ee
This metric covers the right exterior and the white hole interior.

\paragraph{\underline{Regions of Overlap}}

\subparagraph{Right Exterior}

The `Right Exterior' region is described by both the EF1 $(\l, v, w)$
and EF4 $(\l, u_1, \om_1)$ coordinates. These are related by the
following smooth coordinate transformations
\be
v=u_1+\frac1{\sqrt{L}}\log \bigg(\frac{\l-\l_0}{\l+\l_0}\bigg)\hspace{1cm}
w=\om_1+\frac1{\sqrt{\bL}}\log \bigg(\frac{\l-\l_0}{\l+\l_0}\bigg)
\label{rt-exterior}
\ee

\subparagraph{Black Hole Interior}

The `Black Hole Interior' region is described by both the EF1 $(\l, v, w)$ 
and EF2 $(\l_1, u, \om)$ coordinates, which are
related by the following smooth coordinate transformations
\be
v=u+\frac1{\sqrt L}\log\bigg(\frac{\l_0-\l_1}{\l_0+\l_1}\bigg),\hspace{1cm}
w=\om+\frac1{\sqrt{\bL}}\log\bigg(\frac{\l_0-\l_1}{\l_0+\l_1}\bigg),\quad
\l_1= \l 
\ee

\subparagraph{Left Exterior}

The `Left Exterior' region is described by both the EF2 $(\l_1, u, \om)$ 
and EF3 $(\l_1, v_1, \om_1)$ coordinates, which are
related by the following smooth coordinate transformations:
\be
v_1=u+\frac1{\sqrt{L}}\log \bigg(\frac{\l_1-\l_0}{\l_1+\l_0}\bigg)\hspace{1cm}
w_1=\om+\frac1{\sqrt{\bL}}\log \bigg(\frac{\l_1-\l_0}{\l_1+\l_0}\bigg)
\ee

\subparagraph{White Hole Interior}

The `White Hole Interior' finds a description in both the EF3 
$(\l_1, v_1, \om_1)$ and EF4 $(\l, u_1,
\om_1)$ coordinates, which are
related by the following smooth coordinate transformations:
\be
v_1=u_1+\frac1{\sqrt{L}}\log \bigg(\frac{\l_0-\l}{\l_0+\l}\bigg),
\hspace{1cm}
w_1=\om_1+\frac1{\sqrt{\bL}}\log \bigg(\frac{\l_0-\l_1}{\l_0+\l_1}\bigg),
\quad \l= \l_1
\ee

\subsection{Kruskal coordinates}

The union of all the above coordinate patches, together with a
neighbourhood (indicated by K5 in
Fig \ref{fig-ef1234}) of the bifurcation surface (the meeting point of the
past and future horizons in the Penrose diagram)  can be described by a
set of Kruskal coordinates, in which the metric
reads 
\be
ds^2= - \frac1{2\lambda_0} dU dV +\frac1{\sqrt{L}} U dV dy +
\frac{\bL}{4} dy^2
\label{Kruskal}
\ee
The coordinate transformation between various EF coordinates
and the Kruskal coordinates are given below.

\ubsec{1. Right exterior + Black Hole Interior : EF1 to Kruskal}

The transformation from EF1 to the $(U,V,y)$ coordinates is
\be
U=-\exp(-\sqrt{L}v)(\l-\l_0),\hspace{0.2cm}V=\exp(\sqrt{L}v),\hspace{0.2cm}
y=w-\sqrt{\frac{L}{\bL}}v+\frac{1}{\sqrt{\bL}}\log\bigg(\frac{\lambda+\lambda_0}{2\sqrt{\bL}}\bigg)
\label{kruskal-1}
\ee

In the `Right Exterior' region, $\l>\l_0$, while in the `Black Hole Interior', 
$\l<\l_0$. The above transformations give us the metric \eq{Kruskal} in both the
regions.

\ubsec{2. Left Exterior + Black Hole Interior : EF2 to Kruskal}

The transformation from EF2 to $(U,V,y)$ coordinates is
\be
U=\exp(-\sqrt{L}u)(\l_1+\l_0),\hspace{0.2cm}V=-\exp(\sqrt{L}u)\frac{\l_1-\l_0}{\l_1+\l_0},\hspace{0.2cm}
y=\om-\sqrt{\frac{L}{\bL}}u+\frac1{\sqrt{\bL}}\log(\l_1+\l_0)
\label{kruskal-2}
\ee
with,
\be
y_1=y-\frac2{\sqrt{\bL}}\log\bigg(\frac{\l_1+\l_0}{2\sqrt{\bL}}\bigg)
\ee

In the `Black Hole Interior' $\l_1<\l_0$, while in the `Left Exterior' region $\l_1>\l_0$. These coordinate
transformations give us the metric \eq{Kruskal} in both the regions.

\ubsec{3. Left Exterior + White Hole Interior : EF3 to Kruskal}

The transformations from EF3 to the $(U,V,y)$ coordinates is 

\be
U=\exp(-\sqrt{L}v_1)(\l_1-\l_0),\hspace{0.2cm}V=-\exp(\sqrt{L}v_1),\hspace{0.2cm}
y=w_1-\sqrt{\frac{L}{\bL}}v_1+\frac{1}{\sqrt{\bL}}\log\bigg(\frac{\l_1+\l_0}{2\sqrt{\bL}}\bigg)
\label{kruskal-3}
\ee

In the `Left Exterior' region $\l_1>\l_0$, while in the `White Hole Interior',
$\l_1<\l_0$. These transformations give us the metric \eq{Kruskal} in both the 
regions.

\ubsec{4. Right Exterior + White Hole Interior : EF4 to Kruskal}

The transformation from EF4 to the $(U,V,y)$ coordinates is
\be
U=-\exp(-\sqrt{L}u_1)(\l+\l_0),\hspace{0.2cm}V=\exp(\sqrt{L}u_1)\frac{\l-\l_0}{\l+\l_0},\hspace{0.2cm}
y=\om_1-\sqrt{\frac{L}{\bL}}u_1+\frac1{\sqrt{\bL}}\log(\l+\l_0)
\label{kruskal-4}
\ee
with,
\be
y_1=y-\frac2{\sqrt{\bL}}\log\bigg(\frac{\l_1+\l_0}{2\sqrt{\bL}}\bigg)
\ee

In the `White Hole Interior' $\l<\l_0$, while in the `Right Exterior' region $\l>\l_0$. The above
transformations give us the metric \eq{Kruskal} in both the regions. 

\subsection{Poincare\label{sec-poincare}}

In this section we show how the EF1, EF2 coordinates can, in fact, be
obtained from Poincare coordinates $\zeta, X_\pm = X_0 \pm X_1$, in
terms of which the metric is written as
\be
ds^2= \frac1{\zeta^2}(d\zeta^2 - dX_+ dX_-)
\label{poincare}
\ee 
We will choose  $L= \bL$ for simplicity, so $\lambda_0=L/2$. 

The coordinate transformation from $X_\pm, \zeta$ to the EF1 coordinates 
is given by
\begin{eqnarray}
 v&=&\frac{\log(X_+)}{\sqrt{L}},\;
 w=-\frac{1}{\sqrt{L}}\log\left(\frac{-X_+X_-+\zeta^2}{X_+}\right),\;
 \frac{\lambda}{\lambda_0}=\frac{-2X_+ X_- +\zeta^2}{\zeta^2}\
\label{poincare-to-ef1}
\end{eqnarray}
whereas the coordinate transformation 
from $X_\pm, \zeta$ to the EF2 coordinates 
is given by
\begin{eqnarray}
 u&=&\frac{1}{\sqrt{L}}\log\left(\frac{-X_+X_-+\zeta^2}{X_-}\right), \;
 \om =-\frac{\log(X_-)}{\sqrt{L}}, \;
 \frac{\lambda_1}{\lambda_0}=\frac{-2X_+X_-+\zeta^2}{\zeta^2}\
\label{poincare-to-ef2}
\end{eqnarray}
There are similar coordinate transformations between the other charts
EF3/4 and Poincare.\footnote{
\label{single-poincare}As explained in \cite{Hartman:2013qma}
, it is possible to describe the BTZ black string
in terms of a single Poincare chart. The BTZ black {\it hole} is a
quotient of AdS$_3$, which in appropriate coordinates
\cite{Balasubramanian:1998sn} corresponds to the periodic identification of the
spatial direction; the BTZ string discussed in this paper is obtained by
decompactifying the spatial circle, which gives back AdS$_3$.}

\section{The new metrics in the charts EF3 and EF4}

\begin{align}
 \hbox{EF3:}~~ ds^2&=\frac{1}{B^2}
 \left[
 d\lt_1^2+
 A_+^2 d\vt_1^2+
 A_-^2 d\wt_1^2
 +2 A_+ d\ut_1 d\lt_1
 +2 A_-d\wt_1 d\lt_1
 \right. 
 \nonumber\\
 & - \left.
\lt_1 \bigg(
 B^2+2\bigg(
 A_+\frac{H_-''(\wt_1)}{H_-'(\wt_1)}+A_-\frac{H_+''(\vt_1)}{H_+'(\vt_1)}+
 \lt\frac{H_+''(\vt_1)H_-''(\wt_1)}{H_+'(\vt_1)H_-'(\wt_1)}
 \bigg)\bigg)
 d\wt_1 d\vt_1
 \right]
\label{newmetric-3}
\end{align}
where
\begin{eqnarray}
 A_+=\sqrt{L} H_+'(\vt_1)(\lt_1+\lt_{10})- \lt_1 \frac{H_+''(\vt_1)}{H_+'(\vt_1)},\; 
 A_-=\sqrt{\bL} H_-'(\wt_1)(\lt_1+\lt_{10})- \lt_1 \frac{H_-''(\wt_1)}{H_-'(\wt_1)},\;
 B=2(\lt_1+\lt_{10})\nonumber
\end{eqnarray}

\begin{align}
\hbox{EF4}:~~ ds^2&=\frac{1}{B^2}
 \left[
 d\lt^2+
 A_+^2 d\ut_1^2+
 A_-^2 d\omt_1^2
 -2 A_+ d\ut_1 d\lt
 -2 A_-d\omt_1 d\lt
 \right. 
 \nonumber\\
 & - \left.
\lt \bigg(
 B^2-2\bigg(
 A_+\frac{G_-''(\omt_1)}{G_-'(\omt_1)}+A_-\frac{G_+''(\ut_1)}{G_+'(\ut_1)}-
 \lt\frac{G_+''(\ut_1)G_-''(\omt_1)}{G_+'(\ut_1)G_-'(\omt_1)}
 \bigg)\bigg)
 d\omt_1 d\ut_1
 \right]
\label{newmetric-4}
\end{align}
where
\begin{eqnarray}
 A_+=\sqrt{L} G_+'(\ut_1)(\lt+\lt_0)+ \lt \frac{G_+''(\ut_1)}{G_+'(\ut_1)},\; 
 A_-=\sqrt{\bL} G_-'(\omt_1)(\lt+\lt_0)+ \lt \frac{G_-''(\omt_1)}{G_-'(\omt_1)},\;
 B=2(\lt+\lt_0)\nonumber
\end{eqnarray}

\section{UV/IR cutoffs in EF coordinates\label{uv-ir}}

From AdS/CFT it is well-known that in a Fefferman-Graham coordinate
system such as in \eq{banados}, an IR cutoff surface $z=\eps$ in the
asymptotically AdS spacetime corresponds to a UV cutoff $\eps$ in the
CFT. We wish to express the IR cutoff in the geometry in terms of
the EF coordinates. By using the relation 
\be
z = \sqrt{\frac{2}{\lambda_0^2}\bigg(\lambda-\sqrt{\lambda^2-\lambda_0^2}\bigg)}
\ee
we clearly see that  $z=\epsilon$ for $\eps$ small,
corresponds to $\l = 1/\eps^2$.

\section{An alternative to Banados' metric \label{roberts}}

In a beautiful paper \cite{Roberts:2012aq}, Roberts showed that the
Banados metric \eq{banados} can be obtained from the Poincare metric
\eq{poincare} by a Brown-Henneaux type diffeomorphism (an `SGD' in the
language of our paper), given by
\begin{eqnarray}
 X_\pm&=&f_\pm(x_\pm) +\frac{2z^2 f_\pm'(x_\pm)^2 f_\mp''(x_\mp)}{8f_\pm'(x_\pm)f_\mp'(x_\mp)-z^2 f_\pm''(x_\pm)f_\mp''(x_\mp)}\nonumber\\
 \vspace{5 cm}
 \zeta&=&z\  \frac{\left(4f_+'(x_+)f_-'(x_-)\right)^{\frac{3}{2}}}{8f_+'(x_+)
f_-'(x_-)-z^2 f_+''(x_+)f_-''(x_-)} 
\label{roberts-sgd}
\end{eqnarray}
It was shown in \cite{Roberts:2012aq} that the above diffeomorphism
reduces to a \emph{conformal transformation} on the boundary, with the
the following asymptotic form (as z$\rightarrow$0)
\begin{eqnarray}
 X_\pm&=& f_\pm(x_\pm) + O(z^2)\vspace{15 cm}\nonumber\\ \zeta &=&
 z\sqrt{f_+'(x_+)f_-'(x_-)} + O(z^3)
 \label{asymp} 
\end{eqnarray} 
It was also shown in this paper that $L(x_+), \bL(x_-)$ appearing in
\eq{banados} can be obtained from the zero stress tensor through the
conformal transformation $f_\pm$.

\underbar{A different choice of gauge:} The SGD \eq{roberts-sgd} used
by Roberts seems fairly involved compared to the ones we use in this
paper, e.g. \eq{diffeo}. Can we obtain the metric \eq{banados} by a
simpler SGD similar to ours, which nevertheless has the same conformal
asymptotic form \eq{asymp}? The answer turns out to be yes. Indeed the
simplest way of inventing such a transformation is to take the
asymptotic form \eq{asymp} and gauge fix all the higher order terms in
$z$ to 0. We then have a new, exact transformation of the form
\be 
X_\pm= f_\pm(x_\pm), \hspace{1cm} \zeta =
z\sqrt{f_+'(x_+)f_-'(x_-)} 
\label{roberts-new}
\ee
Note the similarity with our SGDs, say
\eq{diffeo} (recall that $z \sim 1/\sqrt\l$ near the boundary).
\eq{roberts-new} transforms the Poincare metric to 
\begin{align}
ds^2=&\frac{dz^2}{z^2}
+\frac{f_+''(x_+)}{z f_+'(x_+)}dx_+dz
+\frac{f_-''(x_-)}{z f_-'(x_-)}dx_-dz
+\frac{1}{4}
\left(
\frac{f_+''(x_+)^2}{f_+'(x_+)^2}dx_+^2
+\frac{f_-''(x_-)^2}{f_-'(x_-)^2}dx_-^2
\right)
\nonumber\\
 & -
\left( 
\frac{2}{z^2}-\frac{f_+''(x_+)f_-''(x_-)}{2f_+'(x_+)f_-'(x_-)}
\right)dx_+ dx_-
\label{roberts-newmetric}
\end{align}
{\it A priori} this is a new metric different from
\eq{banados}. However, the holographic stress tensor
\cite{Balasubramanian:1999re} obtained from this metric is the same as
obtained from \eq{banados} given by \eq{stress}. As discussed in
Section \ref{nontrivial} and \ref{def-nontrivial}, the above metric
and \eq{banados} differ only by a trivial diffeomorphism, and are
hence essentially identical.\footnote{Note that in this new metric
  \eq{roberts-newmetric}, the position of the horizon is at
  $z=\infty$. Of course, it can be brought to a finite value by an
  additional coordinate transformation involving the radial
  coordinate.}  Note that this example shows the enormous gauge
ambiguity in the choice of a metric in AdS$_3$ (whose physical content
is manifested in the boundary behaviour). Indeed, by the same token
even the SGD's employed in this paper are ambiguous; the solutions
presented in Section \ref{behind} are one of a gauge equivalent class
of metrics.


\section{Unitary realization of conformal transformation \label{unitary}}

Under a finite, non-trivial, holomorphic coordinate transformation, $w\rightarrow w'=f(w)$, the stress tensor of a 2D
CFT transforms as
\be
\tilde T(w')=\bigg(\frac{\partial w'}{\partial w}\bigg)^{-2}[T(w)-\frac{c}{12}S(w',w)]
\ee
with the Schwarzian derivative S($w',w$) given by 
\be
S(w',w)=\bigg(\frac{\partial^3 w'}{\partial w^3}\bigg)\bigg(\frac{\partial w'}{\partial w}\bigg)^{-1} 
-\frac{3}{2}\bigg(\frac{\partial^2 w'}{\partial w^2}\bigg)^2\bigg(\frac{\partial w'}{\partial w}\bigg)^{-2}
\label{schwarzian}
\ee
For an infinitesimal transformation $w\rightarrow w'=f(w)=w+\epsilon(w)$, the Schwarzian derivative 
turns out to be
\be
S(w',w)= \epsilon'''(w) + \mathcal{O}(\epsilon^2)
\ee
The change in the stress tensor, under such a transformation, becomes  
\be
\delta T(w)\approx -\eps(w)T'(w)-2\eps'(w)T(w)-\frac{c}{12}\eps'''(w) 
+ \mathcal{O}(\epsilon^2)
\label{tensor}
\ee
Now, the Laurent expansion of $T(w)$ and $\eps(w)$ is
\be
T(w)=\sum_{m=-\infty}^{\infty}\frac{L_m}{w^{m+2}}\hspace{2cm}\eps(w)=\sum_{m=-\infty}^{\infty}\eps_m w^{-m+1}
\label{Laurent}
\ee
where $L^\dagger_n = L_{-n}$, $\eps^\dagger_n= - \eps_{-n}$ and the $L_n$'s 
satisfy the Virasoro algebra
\be
[L_m,L_n]=(m-n)L_{m+n} +\frac{c}{12}m(m^2-1)\delta_{m+n,0}
\label{virasoro}
\ee
Plugging \eq{Laurent} into \eq{tensor}, we get
\be
\delta L_m=\sum_{n=-\infty}^{\infty}\bigg\{
(m+n)L_{m-n}\eps_n+\frac{c}{12}n(n^2-1)\eps_n\delta_{m-n,0}\bigg\}
\label{sten}
\ee
We wish to construct a unitary operator $U=U(\eps)$ which implements
the above conformal transformations, namely that it satisfies  
\be
U(\eps)^\dagger L_m U(\eps)- L_m= \delta L_m + O(\eps^2)
\label{unitary-a}
\ee
The required unitary operator, in fact, is
\be
U(\eps)=\exp(\sum_{n=-\infty}^{\infty} \epsilon_n L_{-n})
\ee
The proof is straightforward. Note that the LHS of
\eq{unitary-a} becomes 
\[
(1- \sum_n \eps_{-n} L_{n}) L_m  (1+ \sum_n \eps_{n} L_{-n}) - L_m
= -\sum_{n=-\infty}^{\infty}\epsilon_{-n}(L_{n}L_m) 
+\sum_{n=-\infty}^{\infty}\epsilon_n(L_m L_{-n})+\mathcal{O}(\eps^2)\  
\]
After flipping the sign of $n$ in the first sum, this becomes 
\[
\eps_n [L_m, L_{-n}]
\]
which  reduces to the expression \eq{sten} upon using the Virasoro
algebra \eq{virasoro}.

Thus, we have explicitly constructed a unitary operator $U$ such
that $U^\dagger T(w) U - T(w)$ is given by \eq{tensor}.

\bibliographystyle{jhepmod}
\bibliography{17_d}
\end{document}